\documentclass[moor]{informs1}              

\usepackage{natbib}
 \NatBibNumeric
 \def\bibfont{\small}%
 \bibpunct[, ]{[}{]}{,}{n}{}{,}%

\usepackage[colorlinks=true,breaklinks=true,bookmarks=true,urlcolor=blue,
     citecolor=blue,linkcolor=blue,bookmarksopen=false,draft=false]{hyperref}

\def\EMAIL#1{\href{mailto:#1}{#1}}
\def\URL#1{\href{#1}{#1}}         

\def\TheoremsNumberedThrough{%
\theoremstyle{TH}%
\newtheorem{Theorem}{Theorem}
\newtheorem{Lemma}{Lemma}
\newtheorem{Proposition}{Proposition}
\newtheorem{Corollary}{Corollary}

\newtheorem{Assumption}{Assumption}
\theoremstyle{EX}
\newtheorem{Remark}{Remark}

\newtheorem{Definition}{Definition}

}

\def\proof{\textit{Proof. }}
\def\qed{\Halmos}

\TheoremsNumberedThrough     

\EquationsNumberedThrough    

\def\be{\begin{eqnarray}}
\def\ee{\end{eqnarray}}

\def\b*{\begin{eqnarray*}}
\def\e*{\end{eqnarray*}}

\newcommand{\rmi}{{\rm (i)$\>\>$}}
\newcommand{\rmii}{{\rm (ii)$\>\>$}}
\newcommand{\rmiii}{{\rm (iii)$\>\>$}}

\def \E{\mathbb{E}}
\def \F{\mathbb{F}}

\def \P{\mathbb{P}}
\def \Q{\mathbb{Q}}
\def \R{\mathbb{R}}

\def\Ac{{\cal A}}
\def\Bc{{\cal B}}

\def\Dc{{\cal D}}
\def\Ec{{\cal E}}
\def\Fc{{\cal F}}
\def\Gc{{\cal G}}
\def\Hc{{\cal H}}

\def\Pc{{\cal P}}
\def\Qc{{\cal Q}}

\def\Uc{{\cal U}}

\def \Om{\Omega}
\def \om{\omega}
\def \Omb{\overline{\Om}}
\def \omb{\bar{\om}}
\def \omh{\hat{\om}}

\def \eps{\varepsilon}

\def \0{\mathbf{0}}
\def \1{\mathbf{1}}
\def \x{\times}

\def \Fcb{\overline{{\cal F}}}
\def \Fbb{\overline{\F}}
\def \Fct{\tilde{{\cal F}}}
\def \Ft{\tilde{\F}}
\def \Pcb{\overline{\Pc}}
\def \Qcb{\overline{\Qc}}

\def \Pb{\overline{\P}}

\def \Hcb{\overline{\Hc}}

\def\Bf{\mathfrak{B}}
\def\Rb{\overline{\R}}

\def\Qb{\overline{\Q}}
\def\Hcb{\overline{\Hc}}

\def\NA2{\mathrm{NA}_2}
\def\NA{\mathrm{NA}}

\def\Omh{\widehat{\Om}}
\def\Pch{\widehat{\Pc}}
\def\Pct{\widetilde{\Pc}}
\def\Qch{\widehat{\Qc}}

\def\Fch{\widehat{\Fc}}
\def\Fbh{\widehat{\F}}
\def\Ph{\widehat{\P}}
\def\Qh{\widehat{\Q}}

\def\CPSt{\mathcal{S}}

\def\scap#1#2{#1\cdot #2}

\begin{document}


 \RUNAUTHOR{Deng, Tan and Yu}

\RUNTITLE{Utility maximization with transaction costs and uncertainty}

\TITLE{Utility maximization with proportional transaction costs under model uncertainty}

\ARTICLEAUTHORS{%
\AUTHOR{Shuoqing Deng}
\AFF{CEREMADE, Universit\'e Paris-Dauphine, PSL  University, Place du Mar{\'e}chal de Lattre de Tassigny, 75016, Paris, France, \EMAIL{shuoqing.deng@gmail.com} \URL{}}
\AUTHOR{Xiaolu Tan}
\AFF{Department of Mathematics, The Chinese University of Hong Kong, Shatin, N.T., Hong Kong, \EMAIL{xiaolu.tan@gmail.com} \URL{}}
\AUTHOR{Xiang Yu}
\AFF{Department of Applied Mathematics, The Hong Kong Polytechnic University, Hung Hom, Kowloon, Hong Kong, \EMAIL{xiang.yu@polyu.edu.hk} \URL{}}
} 

\ABSTRACT{%
We consider a discrete time financial market with proportional transaction costs under model uncertainty,
	and study a num\'eraire-based semi-static utility maximization problem with an
exponential utility preference.
	The randomization techniques recently developed in \cite{BDT17} allow us to transform the original problem into a frictionless counterpart on an enlarged space.
	By suggesting a different
	dynamic programming argument than in
	\cite{bartl2016exponential},
	we are able to prove the existence of the optimal strategy and the convex duality theorem in our context with transaction costs.
	In the frictionless framework, this alternative dynamic programming argument { also} allows us to generalize the main results in \cite{bartl2016exponential} to a weaker market condition.
	Moreover,
	as an application of the duality representation, some basic features of utility indifference prices are investigated in our robust setting with transaction costs. 
}%


\MSCCLASS{Primary: 60G42, 49M29; secondary: 49L20, 91B16.}
\ORMSCLASS{Primary: Probability: Stochastic model applications ; secondary: Finance: Asset Pricing.}
\HISTORY{Received on July 21, 2018; Second version received on February 16, 2019; Accepted on July 21, 2019.}

\maketitle

\section{Introduction}

The optimal investment via utility maximization has always been one of the fundamental problems in quantitative finance. In particular, the optimal semi-static portfolio among risky assets and liquid options and the associated utility indifference pricing of unhedgeable illiquid contingent claims have attracted a lot of research interests recently. In the classical dominated market model, the so-called utility maximization with random endowments was extensively investigated, see among \cite{RasonyiStettner}, \cite{cvitanicutility}, \cite{KramHug}, \cite{Del6authors}, \cite{Bech}, \cite{owenZit}, \cite{BellFri} and \cite{BFG}. In particular, the duality approach has been proposed and developed as a powerful tool to deal with general incomplete market models. Without knowing the specific underlying model structures, the convex duality relationship enables one to obtain the existence of the primal optimizer by solving the corresponding dual optimization problem first. Typically, the dual problem is formulated on the set of equivalent (local) martingale measures (EMM), whose existence is ensured by some appropriate no arbitrage assumptions. Depending on the domain of the utility function, different techniques are involved in order to obtain some convex duality results.
For utilities defined on the positive real line, to handle the random payoffs and to establish the bipolar relationship, the appropriate closure of the dual set of EMM plays the key role, see \cite{cvitanicutility} and \cite{KramHug} for instance. On the other hand, for utilities defined on the whole real line, a subset of EMM with finite general entropy is usually chosen to define the dual problem while the appropriate definition of working portfolios turns out to be critical to guarantee and relate the primal and dual optimizers, see \cite{Del6authors}, \cite{Bech}, \cite{owenZit}, \cite{BellFri} and \cite{BFG} and the references therein.

\vspace{0.5em}

Because of the growing complexity of real financial markets, the aforementioned optimization problems have been actively extended mainly in two directions. The first fruitful extension incorporates the practical trading frictions, namely transaction costs, into decision making and the resulting wealth process. As transaction costs will generically break the (local) martingale property of the self-financing wealth process under EMM, the dual pricing kernel is not expected to be the same as in the frictionless counterpart. Instead, the no-arbitrage condition is closely related to the existence of a pair of dual elements named the consistent price system (CPS). Briefly speaking, the first component of CPS is a process evolving inside the bid-ask spread, while the second component is an equivalent probability measure under which the first component becomes a martingale. However, similar to the case in the frictionless model, for utility maximization with random endowments, the set of CPS can only serve as the first step to formulate the naive dual problem. More efforts are demanded to deal with the random payoffs from options, see some related work in \cite{BenedettiCampi}, \cite{Yuhabit}, \cite{LinYang} and \cite{BayYU}.
\vspace{0.5em}

	The second compelling extension in the literature is to take into account the model uncertainty, for instance the volatility uncertainty, by starting with a set of possibly mutually singular probability measures.
	Namely, different probability measures describe the believes of different investors on the market.
	In the discrete time framework, the no-arbitrage condition and the fundamental theorem in robust finance have been essentially studied in \cite{ABPS, BouchardNutz.13, BFM, BFHMO}, etc. for frictionless markets,
	and in \cite{DolinskySoner.13, BayraktarZhang.13, bouchard2016consistent, Burzoni16, BurzoniSikic} for market with transaction costs.
	 Analogous to the dominated case, the pricing-hedging duality can usually be obtained by studying the superhedging problem under some appropriate no-arbitrage conditions. The non-dominated robust utility maximization in the discrete time frictionless market was first examined by \cite{nutz2016utility},
	where the dynamic programming principle plays the major role to derive the existence of the optimal primal strategy without passing to the dual problem, see some further extensions in \cite{NeufeldSikic, BC, BC2}. In a context where the model uncertainty is represented by a collection of stochastic processes, \cite{RasonyiMeireles} proved the existence of the optimal strategy for the utility function defined either over the positive or over the whole real line. However, whether the convex duality holds remained open in these pioneer work of utility maximization. Recently, \cite{bartl2016exponential} established the duality representation for the exponential utility preference in the frictionless model under some restrictive no arbitrage conditions, which motivates us to reconsider the validity of duality theorem in this paper with proportional transaction costs under weaker market conditions using some distinctive arguments. We also note a recent paper \cite{BartlCK17}, in which the authors proved a robust utility maximization duality using medial limits and a functional version of Choquet's capacitability theorem.

\vspace{0.5em}

	The main objective of this paper is therefore to study the existence of the optimal strategy, the convex duality theorem and the auxiliary dynamic programming principle for a semi-static utility maximization problem with transaction costs in a discrete time framework.
	To be precise, we envision an investor who chooses the optimal semi-static portfolio in stocks and liquid options with an extra random endowment for the case of exponential utility preference and meanwhile each trading incurs proportional transaction fees.
	The core idea of our analysis is to reduce the complexity of transaction costs significantly by employing the randomization method as in \cite{BDT17}. Consequently, the unpleasant mathematical obstacles 
caused by trading fees can be hidden in an enlarged space 
with additional randomness and some techniques in the literature of robust hedging and utility maximization in frictionless models can be modified and adopted. It is worth noting that by applying the randomization approach 
in \cite{BDT17} but with a different and more involved definition of family of probability measures on the enlarged space, \cite{BayBur} recently established
a super-replication duality with transaction cost under a weaker no-arbitrage condition. 

	\vspace{0.5em}

	Our main contributions are the following.
	First, we develop 
	{ a distinctive} dynamic programming argument comparing to \cite{bartl2016exponential} in a frictionless market.
	This allows us to overcome a measurability difficulty in \cite{bartl2016exponential} and hence generalize their main results (duality and existence) under a weaker market condition.
	This generalization is presented in Appendix.
	Secondly, we generalize the randomization technique in \cite{BDT17} in this utility maximization problem,
	which relies essentially on a minimax argument to resolve a filtration enlargement problem.
	While the corresponding convex/concave property is quite natural for the super-replication problem in \cite{BDT17},
	it is much less obvious for the utility maximization problem and we use a log transformation technique in this exponential utility maximization problem.
	Finally, to manifest the value of the duality representation, we also investigate an application to utility indifference pricing. Several fundamental properties of indifference prices including the asymptotic convergence of indifference prices to the superhedging price and some continuity results with respect to random endowments are confirmed in the robust setting with transaction costs.

	\vspace{0.5em}

	The rest of the paper is organized as follows. In Section \ref{sec:main}, we introduce the market model with transaction costs, and show how to reformulate the robust utility maximization problem on a frictionless market on an enlarged space using the randomization method.
	In Section \ref{sec:exp_utility}, we restrict to the case of the exponential utility preference.
	A convex duality theorem and the existence of the optimal trading strategy are first obtained in the presence of both model uncertainty and transaction costs.
	As an application, several properties of the utility indifference prices are concluded.
	Section \ref{sec:proof} mainly provides the proof of the duality result using a dynamic programming argument.

	\paragraph{Notation.} Given a measurable space $(\Om, \Fc)$, we denote by $\Bf(\Om, \Fc)$ the set of all probability measures on $(\Om, \Fc)$.
	For a topological space $\Om$, $\Bc(\Om)$ denotes its Borel $\sigma$-field with the abbreviate notation
	$\Bf(\Om) := \Bf(\Om, \Bc(\Om))$.
	For a Polish space $\Om$, a subset $A \subseteq \Om$ is called analytic if it is the image of a Borel subset of another Polish space under a Borel measurable mapping.
	A function $f: \Om \to \Rb := [-\infty, \infty]$ is upper semianalytic if $\{\om \in \Om ~: f(\om) > a\}$ is analytic for all $a\in \R$.
	Given a probability measure $\P \in \Bf(\Om)$ and a measurable function $f: \Om \to \Rb$, we define the expectation
	$$
		\E^\P[ f] ~:=~ \E^{\P}[f^+] - \E^{\P}[f^-], ~~~\mbox{with the convention}~ \infty - \infty = -\infty.
	$$
	For a family $\Pc \subseteq \Bf(\Om)$ of probability measures, a subset $A \subset \Om$ is called $\Pc$-polar if $A \subset A'$ for some universally measurable set $A'$ satisfying $\P[A'] = 0$ for all $\P \in \Pc$,
	and a property is said to hold $\Pc$-quasi surely or $\Pc$-q.s if it holds true outside a $\Pc$-polar set.
	For $\Q \in \Bf(\Om)$, we write $\Q \lll \Pc$ if there exists $\P' \in \Pc$ such that $\Q \ll \P'$. Given a sigma algebra $\Gc$, we denote by $L^{0}(\Gc)$ the collection of $\R^{d}$-valued random variable that are $\Gc$-measurable, $d$ being given by the context.

\section{Market model and Problem Formulation}
\label{sec:main}
	We first introduce a financial market with proportional transaction costs in a multivariate setting under model uncertainty.
	A utility maximization problem is formulated afterwards and we then reformulate the problem further in a frictionless market setting on an enlarged space.
	Although the reformulation technique can be used for a more general framework, we
	will stay essentially in the context of Bouchard and Nutz \cite{BouchardNutz.13, bouchard2016consistent}.

\subsection{Market model and preliminaries} \label{subsec:prelimin}

\paragraph{A product space with a set of probability measures}
	Let  $\Om_0 := \{\om_0\}$ be a singleton and $\Om_1$ be a Polish space.
	For each $t=1, \cdots, T$, we denote by $\Om_t := \Om_1^t$ the $t$-fold Cartesian product of $\Om_1$ and let $\Fc_t^0 := \Bc(\Om_t)$ and $\Fc_t$ its universal completion.
	In particular, $\Fc_0$ is trivial.
	We define the filtered measurable space $(\Om, \Fc)$ by
	$$
		\Om := \Om_T, \;
		\Fc := \Fc_T,\; \F := (\Fc_t)_{0 \le t \le T}
		~~\mbox{and}~
		\F^0 := (\Fc^0_t)_{0 \le t \le T}.
	$$
	Let us then introduce a set $\Pc$ of probability measures on $(\Om, \Fc)$ by
	\be \label{eq:def_Pc}
		\Pc :=
		\big\{
			\P := \P_0 \otimes \P_1 \otimes \cdots \otimes \P_{T-1} ~: \P_t(\cdot) \in \Pc_t(\cdot) \mbox{ for } t\le T-1
		\big\}.
	\ee
	In the { definition} above, $\P_t:\Omega_{t}\mapsto \Bf(\Omega_{1})$ are probability kernels such that the probability measure $\P$
	is defined by Fubini's theorem in the sense that
	$$
 		\P(A)
		:=
		\int_{\Om_1} \cdots \int_{\Om_1} \mathbf{1}_A (\om_1, \om_2 \cdots, \om_T)
		\P_{T-1}(\om_1, \cdots, \om_{T-1}; d \om_T)  \cdots \P_0(d \om_1),
	$$
	and $\Pc_t(\om) $ is a non-empty convex set in $\Bf(\Omega_{1})$,
	which represents the set of all possible models for the $(t+1)$-th period, given the state $\om \in \Om_t$ at time $t = 0, 1\cdots, T-1$.
	As in the literature, we assume that, for each $t$,
		\be 		
			\label{eq:AnalyticGraph}
			\left[ \left[ \Pc_t \right] \right]
			:=
			\big\{ (\om, \P): \om \in \Om_t,
				\P \in \Pc_t(\om)
			\big\}
			~\subseteq~ \Om_t \times \Pc(\Om_1)
			~~ \text{is analytic.}
		\ee
	This ensures in particular that $\Pc$ in \eqref{eq:def_Pc} is nonempty.
	
\paragraph{A financial market with proportional transaction cost}
	
	The financial market with proportional transaction cost is formulated in terms of random cones.
	Let $d \ge 2$,
	for every $ t \in \{0, 1, \cdots, T\}$, $K_t: \Om \to 2^{\R^{d}}$ is a $\Fc^{0}_t$-measurable random set in the sense that
	$\{\om \in \Om : K_t(\om) \cap O \neq \emptyset \}  \in \Fc^{0}_t$ for every closed (open) set $O \subset \R^{d}$.
	Here, for each $\om \in \Om$, $K_t(\om)$ is a closed convex cone containing $\R^{d}_+$, called the solvency cone at time $t$. It represents the collection of positions, labelled in units of different $d$ financial assets, that can be turned into non-negative ones (component by component) by performing immediately exchanges between the assets.
	We denote by $K^*_t  \subset \R^{d}_+$ its (nonnegative) dual cone:
	\be \label{eq:def_K_star}
		K^*_t(\om) ~:=~ \big \{y \in \R^d  ~: \scap{ x}{y } \ge 0 ~\mbox{for all}~ x \in K_t(\om) \big \},
	\ee
	where $\scap{ x}{y } := \sum_{i=1}^{d} x^i y^i$ is the inner product on $\R^{d}$.
	For later use, let us also  introduce
	$$
		K^{*,0}_t(\om) ~:=~ \big\{ y = (y^1, \cdots, y^d) \in K^*_t(\om), ~y^d = 1\big \}.
	$$
	As in \cite{bouchard2016consistent}, we assume the following conditions throughout the paper:
	\begin{Assumption} \label{assum:1}
		$K^*_t \cap \partial \R^d_+ = \{0 \}$ and $\mathrm{int}K^*_t(\om) \neq \emptyset$ for every $\om \in \Omega$ and $t \le T$.
	\end{Assumption}
	
	It follows from the above assumption and \cite[Lemma A.1]{bouchard2016consistent} that
	$K^*_t$, $K^{*,0}_t$ and $\mathrm{int}K^*_t$ are all $\Fc^0_t$-measurable.
	Moreover, there is a $\F^{0}$-adapted process $S$ satisfying
	\begin{align}\label{eq: S in int}
		S_t(\om) \in K^{*,0}_t(\om)\cap \mathrm{int}K^{*}_t(\om)\;\mbox{ for every $\omega \in \Om$, $t\le T$.}
	\end{align}
	We also assume that transaction costs are bounded and uniformly strictly positive. This is formulated in terms of $S$ above.
	\begin{Assumption}\label{assum:2}
		  There is some constant $c > 1$ such that
		$$
			c^{-1} S^i_t(\om) \le y^i \le c S^i_t(\om),
			~\mbox{for every}~i\le  d-1
			~\mbox{and}~ y   \in K^{*,0}_t(\om).
		$$
	\end{Assumption}
	
		Finally, we define the collection of admissible strategies as follows.
	\begin{Definition} \label{def:admis_stra}
		We say that  an $\F$-adapted process  $\eta = (\eta_t)_{0 \le t \le T}$ is an admissible trading strategy if
		$$
		\eta_t \in - K_t\;\;\;\Pc\mbox{-q.s.} \;\mbox{for all $t\le T$.}
		$$
		We denote by $\Ac$ the collection of all admissible strategies. 	
	\end{Definition}
	The constraint $\eta_{t} \in -K_{t}$ means that $0-\eta_{t}\in K_{t}$, i.e., starting at $t$ with $0$, one can perform immediate transfers to reach the position $\eta_{t}$.
	Then, given $\eta\in \Ac$, the corresponding wealth process associated to a zero initial endowment at time $0$ is
	$\big(\sum_{s=0}^{t} \eta_{s} \big)_{t\le T}$.
	We can refer to \cite{BDT17, bouchard2016consistent} for concrete examples. See also the monograph \cite{KabanovSafarian.09}.

\subsection{A utility maximization problem and its reformulation}
\label{subsec:reform}

 	Let $U : \R \to \R \cup \{-\infty\}$ be a non-decreasing concave utility function.
	We are interested in the following robust utility maximization problem with random endowments:
	 \be\label{eq:def_V0}
		V(\xi)
		~:=~
		\sup_{\eta \in \Ac_0} \inf_{\P \in \Pc}
		\E^{\P} \Big[ U \Big( \Big(\xi + \sum_{t=0}^T \eta_t \Big)^d  \Big) \Big],
	\ee
	where $\Ac_0$ denotes the collection of all $\eta \in \Ac$ such that
	$(\xi + \sum_{t=0}^T \eta_t )^i = 0$ for  $i = 1, \cdots, d-1$.
	
	\begin{Remark}
		Note that \eqref{eq:def_V0} is a num\'eraire based utility maximization problem,
		and the $d$-th asset plays the role of the num\'eraire.
		For an admissible strategy in $\Ac_0$, it is required to liquidate the position of all other assets for $i=1, \cdots, d-1$ at the terminal time $T$.

	\end{Remark}

	The mixture of model uncertainty, transaction costs and random endowments can bring a lot of new mathematical challenges.
	Our paramount remedy to reduce the complexity is to reformulate it on a fictitious market without transaction cost.
	In particular, this allows us to use some well known results and techniques in the existing literature.
		
\paragraph{A frictionless market on the enlarged space}

	Given the constant $c > 1$ in Assumption \ref{assum:2}, we define $\Lambda_1 := [c^{-1}, c]^{d-1}$ and $\Lambda := (\Lambda_1)^{T+1}$, and then introduce the canonical process $\Theta_t(\theta) := \theta_t$, $\forall \theta = (\theta_t)_{t \le T}  \in \Lambda$, as well as the $\sigma$-fields $\Fc_t^{\Lambda} := \sigma (\Theta_s,~ s \le t)$, $t\le T$. We next introduce an enlarged space $\Omb := \Om \x \Lambda$, an enlarged $\sigma$-field $\Fcb:= \Fc \otimes \Fc_T^{\Lambda}$, together with three filtrations $\Fbb^0 = (\Fcb^0_t)_{0 \le t \le T}$, $\Ft = (\Fct_t)_{0 \le t \le T}$ and $\Fbb = (\Fcb_t)_{0 \le t \le T}$ in which $\Fcb^0_t := \Fc^0_t \otimes \Fc_t^{\Lambda}$, $\Fct_t := \Fc_t \otimes \{\emptyset, \Lambda\}$ and $\Fcb_t := \Fc_t \otimes \Fc_t^{\Lambda}$ for $t\le T$.
	
	\vspace{0.5em}
	
	Next, let us introduce our randomized market model with the fictitious underlying stock
	$X = (X_t)_{0 \le t \le T}$ defined by
	\be \label{eq:def_X_t}
		X_t (\omb) := \Pi_{K^{*,0}_t(\om)} [S_t(\om) \theta_t], ~~\mbox{for all}~\omb = (\om, \theta) \in \Omb,\; t\le T,
	\ee
	where $S_t(\om)\theta_t := (S_t^1(\om)\theta_t^1, \cdots, S_t^{d-1}(\om) \theta_t^{d-1}, S_t^d(\om))$,
	and $\Pi_{K^{*,0}_t(\om)}[y]$ stands for the projection of $y \in \R^d$ on the convex closed set $K^{*,0}_t(\om)$.
	It is worth noting that  $S_t\in K^{*,0}_t$ for $t\le T$
	and  that $X$ is $\Fbb^0$-adapted by Lemma 2.6 of \cite{BDT17}.

	\vspace{0.5em}

	We then define two sets of  strategy processes by
	$$
		\Hc := \{ \mbox{All}~ \Ft \mbox{-predictable processes} \}
		~~\mbox{and}~~
		\Hcb := \{ \mbox{All}~ \Fbb \mbox{-predictable processes} \}.
	$$
	Notice that $\Fct_t := \Fc_t \otimes \{\emptyset, \Lambda\}$, and hence a $\Ft$-predictable process can be identified to be a $\F$-predictable process.
	Given a strategy $H \in \Hcb$, the resulting wealth process is given by
	$(H \circ X)_t := \sum_{s = 1}^t \scap{ H_{s}}{( X_s - X_{s-1})} $, $t\le T$.

	\vspace{0.5em}
	
	Finally,  let us introduce some sets of probability measures on the enlarged space $(\Omb, \Fcb)$.
	Let
	$$
		\Pcb ~:=~ \big\{  \Pb\in \Bf(\Omb, \Fcb) ~\mbox{such that}~\Pb|_{\Om} \in \Pc \big\}.
	$$
	We next introduce a subset $\Pcb_{\mathrm{int}} \subset \Pcb $ as follows.
	Recall that $\Omb$ has a product structure as $\Om$.
	More precisely, for a fixed $t \leq T$, let $\Omb_0 := \Om_0 \x \Lambda_1$, $\Omb_t := \Omb_0 \x (\Om_1 \times \Lambda_1)^t$ and
	$\Omb := \Om \x \Lambda = \Omb_T$. For $(\om=(\om_0, \cdots, \om_T ), \theta=(\theta_0, \cdots, \theta_T)) \in \Omb$ and $t \leq T$, we denote $\om^t:= (\om_0, \cdots, \om_t)$, $\theta^t:= (\theta_0, \cdots, \theta_t)$ and $\omb^t:= (\om^t, \theta^t)$.

	\begin{itemize}	
		\item For $t = 0, 1, \cdots, T-1$ and $\omb = (\om, \theta) \in \Omb_t$,
		we define $\Pcb(t,\omb) := \big\{ \Pb \in \Bf(\Om_1 \x \Lambda_1) ~: \Pb|_{\Om_1} \in \Pc_t(\om) \big\}$, and
		\be \label{eq:def_Pcb_t}
			\Pcb_{\mathrm{int}}(t,\omb) := \big\{\Pb \in \Pcb(t,\omb) ~: \delta_{\omb} \otimes \Pb [X_{t+1} \in \mathrm{int} K^*_{t+1}] = 1 \big\},
		\ee
		where $\delta_{\omb} \otimes \Pb$ is a probability measure on $\Omb_{t+1} = \Omb_t \x (\Om_1 \x \Lambda_1)$ and
		$X_{t+1}$ (defined in \eqref{eq:def_X_t}) is considered as a random variable defined on $\Omb_{t+1}$.

		\item Let $\Pcb_{\mathrm{int},\emptyset}$ be the collection of all probability measures $\Pb$ on $\Omb_0$ such that $\Pb[X_0 \in \mathrm{int} K^*_0] = 1$. We define
		$$
			\Pcb_{\mathrm{int}}
			~:=~
			\big\{
				\Pb_{\emptyset} \otimes \Pb_0 \otimes \cdots \otimes \Pb_{T-1} ~:
				\Pb_{\emptyset} \in \Pcb_{\mathrm{int},\emptyset}
				~\mbox{and}~
				\Pb_t(\cdot) \in \Pcb_{\mathrm{int}}(t,\cdot) ~\mbox{for}~ t \le T-1
			\big\},
		$$
		where $\Pb_t(\cdot)$ is a universally measurable selector of $\Pcb_{\mathrm{int}}(t,\cdot)$.
	\end{itemize}

	\begin{Remark}\label{remark:panalytic}
		Assume that the analyticity condition \eqref{eq:AnalyticGraph} for $\left[ \left[ \Pc_t \right] \right]$ holds, Lemma 2.13 of \cite{BDT17} asserts that
		\b*
			\left[ \left[ \Pcb_{\mathrm{int}}(t) \right] \right]
			:=
			\big\{ (\omb, \Pb): \omb \in \Omb_t,
				\Pb \in \Pcb_{\mathrm{int}}(t,\omb)
			\big\}
			~\text{is also analytic,}
		\e*
		which in particular ensures that $\Pcb_{\mathrm{int}}$ is nonempty.
	\end{Remark}

\vspace{0.5em}


\paragraph{Reformulation on the enlarged space}

	We now reformulate the utility maximization \eqref{eq:def_V0} on the enlarged space $\Omb$ { using the} underlying stock $X$.
	Let us set
	$$
		g(\omb) := \scap{ \xi(\om) }{ X_T(\omb) },
		~\mbox{for all}~
		\omb = (\om, \theta) \in \Omb,
	$$
	as the contingent claim.

	\begin{Proposition} \label{prop:reformulation}
		Suppose that Assumptions \ref{assum:1} and \ref{assum:2} hold, then
		$$
			 V(\xi)
			~=~
			\sup_{H \in \Hc} \inf_{\Pb \in \Pcb} \E^{\Pb}
			\Big[ U \Big( g + (H \circ X)_T \Big) \Big]
			~=~
			\sup_{H \in \Hc} \inf_{\Pb \in \Pcb_{\mathrm{int}}} \E^{\Pb}
			\Big[ U \Big( g + (H \circ X)_T \Big) \Big].
		$$
	\end{Proposition}
	\proof To simplify the notation, let us write $\Delta X_{t}:=X_{t}-X_{t-1}$. We shall follow closely the arguments in Proposition 3.3 of \cite{BDT17}.
	
	\vspace{0.5em}
	
	\noindent \textit{Step 1}: Fix $\eta \in \Ac_0$ and define the $\Ft$-predictable process $H$ by $H_t := \sum_{s=1}^t \Delta H_s$ with $\Delta H_t := \eta_{t-1}$ for $t=1, \cdots, T$.
	By rearranging all terms, we have
	$$
		\left(\xi + \sum_{t=0}^T \eta_t \right)^d
		=
		\scap{\left( \xi + \sum_{t=0}^T \eta_t \right)}{ X_T }
		=
		\sum_{t=1}^T H_t \cdot \Delta X_t + \sum_{t=0}^T \scap{ \eta_t}{ X_t } + g
		~\le~
		\sum_{t=1}^T H_t\cdot  \Delta X_t + g,
	$$
	where the last inequality follows by the fact that $\eta_t \in -K_t$ and hence $\eta_t \cdot X_t \le 0$.
	As $U$ is non-decreasing, it follows that
	$$
		\inf_{\P \in \Pc} \E^{\P} \Big[ U \Big( \Big(\xi + \sum_{t=0}^T \eta_t \Big)^d  \Big) \Big]
		~\le~
		\inf_{\Pb \in \Pcb} \E^{\Pb} \Big[ U \Big( g + (H \circ X)_T \Big) \Big],
	$$
	which yields that
	$$
		 V(\xi) ~\le~ \sup_{H \in \Hc} \inf_{\Pb \in \Pcb} \E^{\Pb} \Big[ U \Big( g + (H \circ X)_T \Big) \Big].
	$$
	{By the same argument using $\Pcb_{\mathrm{int}}$ to replace $\Pcb$, we can similarly obtain the inequality}
	$$
		 V(\xi) ~\le~ \sup_{H \in \Hc} \inf_{\Pb \in \Pcb_{\mathrm{int}}} \E^{\Pb} \Big[ U \Big( g + (H \circ X)_T \Big) \Big].
	$$

	\noindent \textit{Step 2}: To prove the reverse inequality, we fix $H \in \Hc$.
	Define $\eta = (\eta_t)_{0 \le t \le T}$ by
	$\eta^i_t :=\Delta H^i_{t+1}$, $t\le  T-1$
	and $\eta^i_T := - \xi^i - \sum_{s=0}^{T-1} \eta^i_s$ for $i\le d-1$,
	and
	\be \label{eq:H2eta}
		\eta^d_t(\om) := \inf_{\theta \in \Lambda} m^d_t(\om, \theta)
		~\mbox{with}~ m^d_t(\omb) := - \sum_{i=1}^{d-1} \eta^i_t(\om) X^i_t(\omb),~
		t\le  T.
	\ee
	for all $\omb = (\om, \theta) \in \Omb$.
	{ As} $m^d_t(\om, \theta)$ is bounded and continuous in $\theta$, $\eta^d_t $ is  $\Fc_t$-measurable by the Measurable Maximum Theorem (see e.g. Theorem 18.19 of \cite{AliprantisBorder}).
	From the construction, we know $\eta \in \Ac_0$.
	Thus we have
	\begin{align}
	\inf_{\theta \in \Lambda }  \big( (H \circ X)_T + g \big) (\cdot, \theta)
	& =   \inf_{\theta \in \Lambda }  \Big \{   \scap{ \Big( \sum_{t=0}^{T} \eta_t+\xi \Big)}{ X_T  }- \sum_{t=0}^{T-1} \scap{ \eta_t}{  X_{t} }    \Big\} (\cdot, \theta)  \nonumber \\
	& =   \inf_{\theta \in \Lambda }  \Big\{\scap{ \Big(\sum_{t=0}^{T} \eta_t+\xi \Big)}{ X_T  }\Big\} (\cdot, \theta)- \sum_{t=0}^{T-1} \sup_{\theta \in \Lambda} \Big\{\scap{ \eta_t}{ X_{t} } \Big\} (\cdot, \theta). \nonumber \\
	& = \inf_{\theta \in \Lambda }  \Big\{\scap{ \Big(\sum_{t=0}^{T} \eta_t+\xi \Big)}{ X_T  }\Big\} (\cdot, \theta)
	~=~
	\Big(\xi + \sum_{t=0}^T \eta_t \Big)^d,
	\end{align}
	where in the second equality we exchange the the infimum and the summation, because each $X_t$ depends on $\theta$ only through $\theta_t$ for $t=0, \cdots, T$.
	Let $\eps > 0$, we can use a measurable selection argument (see e.g. Proposition 7.50 of \cite{BertsekasShreve.78}) to choose a
	universally measurable map $\om \in \Om \mapsto \theta_{\eps}(\om) \in \Lambda$ such that, for all $\om \in \Om$, one has
	$$
		U\Big( \big(H \circ X \big)_T \big(\om, \theta_{\eps}(\om) \big) + g(\om, \theta_{\eps}(\om)) \Big)
		~\le~
		\inf_{\theta \in \Lambda} U\Big( \big(H \circ X \big)_T(\om, \theta) + g(\om, \theta) \Big) + \eps,
	$$
	where the r.h.s. term is a universally measurable random variable defined on $\Om$.
	Then given $\P \in \Pc$, one defines $\Pb_{\eps} := \P \circ (\om, \theta_{\eps}(\om))^{-1} \in \Pcb$ and obtains
	$$
		\E^{\Pb_{\eps}} \Big[ U \big((H \circ X)_T + g \big) \Big]
		~\le~
		\E^{\P} \Big[\inf_{\theta \in \Lambda} U\Big( \big(H \circ X \big)_T(\cdot, \theta) + g(\cdot, \theta) \Big) \Big]
		~+~ \eps.
	$$
	By arbitrariness of $\eps >0$ and the fact that $\Pb_{\eps} \in \Pcb$, it follows that
	\begin{eqnarray} \label{eq:inf theta pour uti}
		\inf_{\Pb \in \Pcb} \E^{\Pb} \Big[ U \big((H \circ X)_T + g \big) \Big]
		&\le&
		\inf_{\P \in \Pc}  \E^{\P} \Big[  \inf_{\theta \in \Lambda} U\Big( (H \circ X)_T(\cdot, \theta) + g(\cdot, \theta) \Big) \Big] \\
		&=&
		\inf_{\P \in \Pc}  \E^{\P} \Big[ U\Big( \inf_{\theta \in \Lambda} \big[  (H \circ X)_T(\cdot, \theta) + g(\cdot, \theta) \big]\Big) \Big] \nonumber \\
		&=&
		\inf_{\P \in \Pc} \E^{\P} \Big[ U \Big( \Big(\xi + \sum_{t=0}^T \eta_t \Big)^d  \Big) \Big]. \nonumber
	\end{eqnarray}

	This leads to
	$$
		\sup_{H \in \Hc} \inf_{\Pb \in \Pcb} \E^{\Pb} \Big[ U \Big( g + (H \circ X)_T \Big) \Big]
		~\le~ V(\xi),
	$$
	and hence we have the { desired} equality.

	\vspace{0.5em}
	
	\noindent \textit{Step 3}:
	For the case with $\Pcb_{\mathrm{int}}$ in place of $\Pcb$,
	it is enough to notice as in \textit{Step 2} that
	$$
		{ \inf_{\theta \in \Lambda_{\mathrm{int}}(\cdot)} \big[  (H \circ X)_T(\cdot, \theta) + g(\cdot, \theta) \big]
		 =
		 \Big(\xi + \sum_{t=0}^T \eta_t \Big)^d ,}
	$$
	where $\Lambda_{{\rm int}}(\omega)$ is
	defined as the collection of $\theta \in \Lambda$ such that $S_{t}(\omega)\theta_{t}\in {\rm int}K^{*}_{t}(\omega)$.
	
	Next, for each $\theta \in \Lambda$,
	we define $A^{\theta}_{t}(\om):=\emptyset$ for $s\ne t$ and $A^{\theta}_{t}(\om):=\{ S_{t}(\om) \theta_{t}\in {\rm int}K^{*}_{t}(\om) \}$.
	Note that $\om \mapsto \mbox{int}K^*_t(\om)$ is $\Fc_t^0$-measurable. Then $\big\{(\om, y) \in \Om_t \x \R ~: S_t(\om) \theta_t = y ~\mbox{and}~ y \in \mbox{int}K^*_t(\om) \big\}$ is a Borel set and hence $\om \mapsto 1_{A^{\theta}_t(\om)}$ is a universally measurable map.
	We then define the universally measurable probability kernels by
	\be \label{eq:kernel_q_theta}
		q^{\theta}_{t}: \om \in \Omega \mapsto  q^{\theta}_{t}(\cdot|\om)
		:=
		\delta_{\theta_{t}} 1_{A^{\theta}_{t}(\om)}
		+\delta_{{\1}}1_{(A^{\theta}_{t}(\om))^{c}}\in \Bf(\Lambda_{1}),~ t\le T,
	\ee
	where  ${\bf 1}$ is the vector of $\R^{d}$ with all entries equal to $1$,
	and $\Bf(\Lambda_{1})$ denotes the collection of all Borel probability measures on $\Lambda_1$.

	\vspace{0.5em}

	It follows that $\P\otimes (q^{\theta}_{0}\otimes q^{\theta}_{1}\otimes\cdots\otimes q^{\theta}_{T}) \in \Pcb_{\mathrm{int}}$ for every $\P \in \Pc$. Then it suffices to argue as in \textit{Step 2} above to obtain that
	\b*
	\inf_{\Pb \in \Pcb_{\mathrm{int}}} \E^{\Pb} \Big[ U \Big( g + (H \circ X)_T \Big) \Big]
	&\le& \inf_{\P \in \Pc}  \E^{\P} \left[   U\Big( \inf_{\theta \in \Lambda_{{\rm int}}(\cdot)} \Big( (H \circ X)_T(\cdot, \theta) + g(\cdot, \theta) \Big) \Big)\right] \\
	&=& 	\inf_{\P \in \Pc}  \E^{\P} \Big[  U\Big(  \Big(\xi + \sum_{t=0}^T \eta_t \Big)^d  \Big) \Big],
	\e*
	and we hence conclude as in \textit{Step 2}.
	\qed

\subsection{The robust no-arbitrage condition of Bouchard and Nutz}

	To conclude, we will discuss the no-arbitrage condition on $\Om$ and its link to that on the enlarged space $\Omb$.
	
	\begin{Definition}
		\rmi We say the robust no-arbitrage condition of second kind $\NA2(\Pc)$ on $\Om$ holds true
		if for all $t \le  T-1 $ and all $\xi \in L^0(\Fc_t)$,
		\b*
			\xi \in K_{t+1} ~~ \Pc \mbox{-q.s.} ~~~ \mbox{implies} ~~~ \xi \in K_{t} ~~ \Pc \mbox{-q.s.}
		\e*

		\vspace{0.5em}
		
		\noindent \rmii Let $(\Q, Z)$ be a couple where $\Q \in \Bf(\Om)$ and $Z = (Z_t)_{t=0, \cdots, T}$ an adapted process,
		$(\Q, Z)$ is called a strict consistent price system (SCPS) if
		$\Q \lll \Pc$, $Z_t \in  \mathrm{int} K_t^*$ $\Q$-a.s. for all $t = 0,\cdots, T$ and $Z$ is a $\Q$-martingale.
	\end{Definition}
	We denote by $\CPSt$ the collection of all SCPS, and also denote the subset
	\be \label{eq:CPS0}
		\CPSt_0
		~:=
		\big\{
		(\Q, Z)\in \CPSt
		~\mbox{such that}~
		Z^d \equiv 1
		\big\}.
	\ee
	\begin{Remark}
		As stated in the fundamental theorem of asset pricing proved in \cite{bouchard2016consistent} (see also \cite{BayraktarZhang.13, BayraktarZhangZhou}),
		the no-arbitrage condition $\NA2(\Pc)$ is equivalent to:
		for all $t \le  T-1 $, $\P \in \Pc$ and $\Fc_t$-random variable $Y$ taking value in $\mathrm{int}K_t^{*}$,
		there exists a SCPS $(\Q, Z)$ such that $\P \ll \Q$, $\P=\Q$ on $\Fc_t$ and $Y=Z_t ~~ \P\mbox{-a.s.}$.
	\end{Remark}
	
	On the enlarged space $\Omb$, we also follow \cite{BouchardNutz.13} to introduce a notion of the robust no-arbitrage condition.

	\begin{Definition} \label{def:NA}
		We say that the robust no-arbitrage condition $\NA(\Pcb_{\mathrm{int}})$ on $\Omb$ holds true if, for every  $H\in  \Hcb$,
		$$
			(H \circ X)_T \ge 0,~ \Pcb_{\mathrm{int}}\mbox{-q.s.}
			~~~\Longrightarrow~~~
			(H \circ X)_T = 0,~ \Pcb_{\mathrm{int}}\mbox{-q.s.}
		$$
	\end{Definition}
	
	\begin{Remark}
		The fundamental theorem of asset pricing in \cite{BouchardNutz.13} proves that
		the condition $\NA(\Pcb_{\mathrm{int}})$ (resp. $\NA(\Pcb)$ )
		is equivalent to :
		for all  $\Pb \in \Pcb_{\mathrm{int}}$ (resp. $\Pcb$ ), there exists $\Qb \in \Bf(\Omb)$ such that $\Pb \ll\Qb \lll \Pcb_{\mathrm{int}}$ (resp. $\Pcb$ ) and $X$ is an $(\Fbb, \Qb)$-martingale.
	\end{Remark}
	
	Hereafter, we denote by $\Qcb_{0}$ the collection of measures $\Qb \in \Bf(\Omb)$ such that $\Qb \lll \Pcb_{\mathrm{int}}$ and $X$ is an $(\Fbb, \Qb)$-martingale. The above two no-arbitrage conditions on $\Om$ and on $\Omb$ are related by Proposition 2.16 of \cite{BDT17}, that we recall as below.
	\begin{Proposition} \label{prop:equiv_NA}		
		The condition $\NA2(\Pc)$ on $\Om$ is equivalent to the condition $\NA(\Pcb_{\mathrm{int}})$ on $\Omb$.
	\end{Proposition}

\section{Exponential utility maximization}		
\label{sec:exp_utility}
	
	Starting from this section, we will restrict ourselves to the case of the exponential utility function, i.e.,
	$$
		U(x):=-\mbox{exp}(-\gamma x),
		~~\mbox{for some constant}~\gamma > 0,
	$$
	and provide a detailed study on the corresponding utility maximization problem.
	
	\vspace{0.5em}
	
	We will consider a general context,
	where one is allowed to trade some liquid options statically at the initial time whose payoffs would also contribute to the terminal wealth.
	Namely, for $e\in\mathbb{N}\cup \{0\}$,
	there are a finite class of $\Fc^0_T$-measurable random vectors $\zeta_i : \Om \to \R^d$, $i=1, \cdots, e$,
	where each $\zeta_i$ represents the payoff of some option $i$ labeled in units of $d$ risky assets.	
	Let $\xi: \Om \to \R^{d}$ represent the payoff of the random endowment,
	then our maximization problem is given by.
	 \be \label{valprim-prob}
		V(\xi, \gamma)
		~:=~
		\sup_{(\ell,\eta) \in {\Ac_e}} \inf_{\P \in \Pc}
		\E^{\P} \left[ U \Big( \Big(\xi + \sum_{i=1}^e \big(\ell_i  \zeta_i - |\ell_{i}| c_{i} \1_{d} \big) + \sum_{t=0}^T \eta_t \Big)^d  \Big) \right].
	\ee
	 where $\1_{d}$ is the vector with all components equal to $0$ but the last one that is equal to $1$ and $\Ac_e$ denotes the collection of all $(l,\eta) \in \R^e \times \Ac$
	 such that $\xi + \sum_{i=1}^e \big(\ell_i  \zeta_i - |\ell_{i}| c_{i} \1_{d} \big) + \sum_{t=0}^T \eta_t \Big)^i=0$ for $i=1, \cdots, d-1$.
	In above, we write $\gamma$ in $V(\xi, \gamma)$ to emphasize the dependence of value in parameter $\gamma$ in the utility function $U$.
	Also, each static option $\zeta_i$ has price $0$, but the static trading induces the proportional transaction cost with rate $c_i > 0$.
	
\subsection{The convex duality result}

	In the robust frictionless setting, 
	the same exponential utility maximization problem has been studied by Bartl \cite{bartl2016exponential},
	in which a convex duality theorem has been established.
	Here, we apply and generalize their results in our context with transaction costs under weaker market conditions.
	
	Let us introduce a robust version of the relative entropy associated to a probability measure $\Q$ as
	\be \label{eq:def_entropy}
		\Ec(\Q,\Pc):=\inf_{\P \in \Pc} \Ec(\Q,\P),
		~~\mbox{where}~
		\Ec(\Q,\P):=\left\{
		\begin{array}{lcl}
			\E^{\P}\left[\frac{d\Q}{d\P}\mbox{log}\frac{d\Q}{d\P}\right],
				&& \mbox{if}~\Q \ll \P, \\
			+\infty,
				&& \mbox{otherwise}.
		\end{array}
		\right.
	\ee
	Note that $\CPSt_0$ is a subset of the collection of SCPS $(\Q,Z)$ defined in \eqref{eq:CPS0}, we then define
	\b*
		 \CPSt^*_e
		:=
		\Big\{ (\Q,Z) \in \CPSt_0
		~:
		\E^{\Q} \big[(\xi \cdot Z_T)_- \big]
		\!+\!
		\Ec(\Q,\Pc) < + \infty
		~\mbox{and}~
		\E^{\Q} \big[\zeta_i \cdot Z_T \big] { \in [-c_i, c_i]}, ~i=1,\! \cdots\!, e
		\Big\}.
	\e*
	

        \begin{Theorem} \label{thm:utility_max_duality}
	Let $\xi$ and $(\zeta_i)_{i\le e}$$: \Om \to \R^d$ be Borel measurable and assume that $\NA2(\Pc)$ holds. Assume either that $e=0$,
		or that $e \ge 1$  and for all $\ell \in \R^{e}$ and $\eta\in \Ac$,
		\be \label{eq: NA avec option et CT}
			\sum_{i=1}^{e}   \big(\ell_i  \zeta_i - |\ell_{i}| c_{i}  \1_{d} \big)+\sum_{t=0}^{T} \eta_{t}\in K_{T}~\Pc\mbox{-q.s.}
			~~\Longrightarrow~~
			\ell =0.
		\ee
		Then, we have
		\be \label{eq:util_max_dual}
			V(\xi, \gamma)
			~=~
			- \exp\Big(
				- \inf_{ (\Q,Z) \in \CPSt_e^*}
				\!\!
				\big\{ \E^{\Q} \big[ \gamma \xi \cdot Z_T\big] + \Ec(\Q,\Pc) \big \}
			\Big),
		\ee
		Moreover, the infimum over $(\ell, \eta)\in\mathcal{A}_e$ is attained by an optimal strategy $(\hat \ell, \hat \eta)$.
	\end{Theorem}

	\begin{Remark}
		Note that up to taking logarithm on both sides and replacing $\gamma \xi$ by $-\xi$, the equality \eqref{eq:util_max_dual} is equivalent to
		\be \label{eq:util_max_dual_reform1}
		\begin{split}
			&\inf_{(\ell,\eta) \in \Ac_e}
			\sup_{\P \in \Pc}
			\log \E^{\P} \left[ \exp \left( \left(  \xi - \sum_{i=1}^{e}   \left(\ell_i  \zeta_i - |\ell_{i}| c_{i} \1_{d} \right) 			- \sum_{t=0}^T \eta_t \right)^d  \right) \right] \\
			=&\!\!\sup_{ (\Q,Z) \in \CPSt_e^*} \left\{ \E^{\Q} \big[ \xi \cdot Z_T\big] - \Ec(\Q,\Pc) \right \}.
		\end{split}
	        \ee
	\end{Remark}

	\begin{Remark}
		When $e \ge 1$,
		$\zeta_i$ is considered as statically traded options and $c_i >0$ is the corresponding proportional transaction cost,
		then the condition \eqref{eq: NA avec option et CT} should be understood as a kind of robust no-arbitrage condition as defined in \cite{BouchardNutz.13}.
		For simplicity, let us consider the case $e=1$.
		By 
		following 
		arguments 
		in Proposition 3.3 of \cite{BDT17},
		$\ell_1  \zeta_1 - |\ell_1| c_1  \1_{d} +\sum_{t=0}^{T} \eta_{t}\in K_{T}~\Pc\mbox{-q.s.}$
		can be shown as equivalent to
		$$
			\ell_1 g_1\big(\omb, \widehat \theta\big)
			+
			\Big(\sum_{t=1}^T H_t \Delta X_t \Big)(\omb) \ge 0, ~\Pcb \mbox{-q.s. and for both}~ \widehat \theta=\pm 1,
		$$
		where $H_t := \sum_{s=0}^{t-1} \eta_s$ and $g_1(\omb, \pm 1) := \zeta_1 \cdot X_T \pm c_1$.
		The robust no-arbitrage condition in Definition \ref{def:NA} will lead to
		$$
			\ell_1 g_1\big(\omb, \widehat \theta\big)
			+
			\Big(\sum_{t=1}^T H_t \Delta X_t \Big)(\omb) = 0, ~\Pcb \mbox{-q.s. and for both}~ \widehat \theta=\pm 1.
		$$
		As $g_1(\omb, 1) \neq g_1(\om, -1)$ when $c_1 > 0$, one obtains $\ell_1 = 0$.
		
	\end{Remark}

	\begin{Remark}
		\rmi The existence of optimal trading strategy $(\hat \ell, \hat \eta)$ in Theorem \ref{thm:utility_max_duality} is an auxiliary result in the proof of duality \eqref{eq:util_max_dual} in our context with exponential utility function $U(x) := - \exp(- \gamma x)$.
		Both duality and the existence of optimal strategy rely crucially on the minimax argument (Lemma \ref{lemm:utility_max_selec_h}) which uses the affine feature of the exponential utility.

		\vspace{0.5em}

		\noindent \rmii
		In the robust context and for general utility functions (with or without transaction cost),
		different results on the existence of the optimal strategy have been obtained in the literature.
		Nutz \cite{nutz2016utility} seems to be the first to introduce this discrete time robust utility maximization problem and obtains the existence result for general utility functions bounded from above and defined on the positive real line.
		Blanchard and Carassus \cite{BC} were able to relax the boundedness condition to some integrability condition.
		Neufeld and Sikic \cite{NeufeldSikic} study the robust utility maximization problem with friction and obtain some existence result under a linear type of no-arbitrage condition.
		Rasonyi and Meireles-Rodrigues \cite{RasonyiMeireles} use a Koml{\'o}s-type argument to prove the existence of the optimal strategy.
		Bartl et al. \cite{BartlCK17} study similar problem by the medial limit argument.

		\vspace{0.5em}
		\noindent \rmiii After the completion of our paper,
		Bayraktar and Burzoni \cite{BayBur} provided a generalization of the randomization approach in \cite{BDT17}
		and proved a pricing-hedging duality under a weaker no-arbitrage condition than the $\NA2(\Pc)$ condition.
		Their generalized randomization approach should also allow to study the above utility maximization problem under the weak no-arbitrage condition.
	\end{Remark}

\subsection{Properties of utility indifference prices}
	It is well known that the superhedging price is too high in practice.
	As an alternative way, the utility-based indifference price has been actively studied,
	in which the investor's risk aversion is inherently incorporated.
	This section presents an application of the convex duality relationship $\eqref{eq:util_max_dual_reform1}$ for the exponential utility maximization and provides some interesting features of indifference prices in the presence of both proportional transaction costs and model uncertainty. Generally speaking, the indifference pricing in our setting can be generated by semi-static trading strategies on risky assets and liquid options.

	\vspace{0.5em}

In the robust framework, similar to Theorem $2.4$ of \cite{bartl2016exponential} in the frictionless model, the duality representation \eqref{eq:util_max_dual_reform1} can help us to derive that the asymptotic indifference prices converge to the superhedging price as the risk aversion $\gamma\rightarrow\infty$ regardless of the transaction costs. To see this, let us first recall the superhedging price defined by
	\be
		\begin{split}
		\pi(\xi)~&:=\inf\left\{y+\sum_{i=1}^{e}c_i|\ell_i|:y \1_d+\sum_{i=1}^{e}\ell_i\zeta_i+\sum_{t=0}^{T}\eta_t-\xi\in
			K_T,\mathcal{P}-q.s., (\ell,\eta)\in\mathcal{A}_e\right\}\\
		&=\sup_{(\mathbb{Q},Z)\in \mathcal{S}_e}\mathbb{E}^{\mathbb{Q}}[\xi\cdot Z_T],\nonumber
		\end{split}
	\ee
	where the equality follows from Theorem $3.1$ of \cite{BDT17} with
	\b*
		\CPSt_e
		:=
		\Big\{ (\Q,Z) \in \CPSt_0
		~:
		\E^{\Q} \big[(\zeta_i \cdot Z_T) \big] { \in [-c_i, c_i]}, ~i=1, \cdots, e
		\Big\}.
	\e*

	The indifference price $\pi_{\gamma}(\xi) \in \R$ of derivative option $\xi$ is, one the other hand, defined by equation
	\be \label{indiffdefinition}
		V(0 \1_d, ~\gamma)
		~=~
		V \big( \pi_{\gamma}(\xi) \1_d  -\xi , ~\gamma \big),
	\ee
	where $V(\cdot)$ is defined by \eqref{valprim-prob}.
	Plugging the expression of $V(\cdot)$ into \eqref{indiffdefinition},
	and recall that $U(x) = - e^{-\gamma x}$, we obtain
	\be
		\exp(- \gamma \pi_{\gamma}(\xi)) &\times& \sup_{(\ell,\eta)\in \mathcal{A}_e}\inf_{\P \in \Pc}
			 \E^{\P} \Big[ - \exp \Big(-\gamma\Big(-\xi + \sum_{i=1}^{e}   \left(\ell_i  \zeta_i - |\ell_{i}| c_{i} \1_{d} \right) + \sum_{t=0}^T \eta_t \Big)^d  \Big) \Big]  \nonumber \\
		&=&~~\sup_{(\ell,\eta)\in \mathcal{A}_e}\inf_{\P \in \Pc}
			 \E^{\P} \Big[ - \exp \Big(-\gamma\Big( \sum_{i=1}^{e}   \left(\ell_i  \zeta_i - |\ell_{i}| c_{i} \1_{d} \right) + \sum_{t=0}^T \eta_t \Big)^d  \Big) \Big].
	\ee
	By the duality representation \eqref{eq:util_max_dual}, we finally have that
	\be\label{indifftwoequiv}
		\pi_{\gamma}(\xi)
		= \sup_{ (\Q,Z) \in \CPSt_e^*} \left\{ \E^{\Q} \big[ \xi \cdot Z_T\big] -\frac{1}{\gamma} \Ec(\Q,\Pc)\right\}  - \sup_{ (\Q,Z) \in \CPSt_e^*} \left\{-\frac{1}{\gamma} \Ec(\Q,\Pc)\right\}.
	\ee

The formula $(\ref{indifftwoequiv})$ yields directly the next few properties of the utility indifference price.
\begin{Lemma}\label{basicpropind}
The following basic properties hold:
\begin{itemize}
\item[$\mathrm{(i)}$] $\pi_{\gamma}(\xi)$ does not depend on the initial wealth $x_0$.
\item[$\mathrm{(ii)}$] $\pi_{\gamma}(\xi)$ is increasing in $\gamma$ (monotonicity in $\gamma$).
\item[$\mathrm{(iii)}$] $\pi_{\gamma}(\beta\xi)=\beta\pi_{\beta\gamma}(\xi)$ for any $\beta\in(0,1]$ (volume scaling).
\item[$\mathrm{(iv)}$] $\pi_{\gamma}(\xi+c)=c+\pi_{\gamma}(\xi)$ for $c\in\mathbb{R}$ (translation invariance).
\item[$\mathrm{(v)}$] $\pi_{\gamma}(\alpha\xi_1+(1-\alpha)\xi_2)\leq \alpha\pi_{\gamma}(\xi_1)+(1-\alpha)\pi_{\gamma}(\xi_2)$ (convexity).
\item[$\mathrm{(vi)}$] $\pi_{\gamma}(\xi_1)\leq \pi_{\gamma}(\xi_2)$ if $\xi_1\leq \xi_2$ (monotonicity).
\end{itemize}
\end{Lemma}

The next result shows the risk-averse asymptotics on the utility indifference prices. Similar results can also be found in \cite{CarRasy, BC2, bartl2016exponential}.
\begin{Proposition}\label{prop:utilityindiff}
In the robust setting of Theorem \ref{thm:utility_max_duality} with proportional transaction costs, we have
\be\label{convindfftosupdg}
\pi(\xi)=\lim_{\gamma\rightarrow\infty}\pi_{\gamma}(\xi).
\ee
\end{Proposition}

We postpone the proof of the above result to Section \ref{sec:utilityindiff}, as it demands some notations and results given afterwards.


\begin{Remark}
Observing the scaling property in item $(iii)$ of Lemma $\ref{basicpropind}$, the limit $(\ref{convindfftosupdg})$ can be rewritten as $\lim_{\beta\rightarrow\infty}\frac{1}{\beta}\pi_{\gamma}(\beta\xi)=\pi(\xi)$, in which the term $\frac{1}{\beta}\pi_{\gamma}(\beta\xi)$ can be understood as the price per unit for a given amount volume $\beta$ of the contingent claim $\xi$.
\end{Remark}

Furthermore, with increasing risk aversion, the convex duality result \eqref{eq:util_max_dual_reform1} also yields that the optimal hedging strategies under the exponential utility preference converge to the superhedging counterpart in the following sense.
\begin{Proposition}
We have that
\be
\lim_{\gamma\rightarrow\infty}\sup_{\mathbb{P}\in\mathcal{P}}\mathbb{E}^{\mathbb{P}}\left[\left( \pi(\xi)\1_{d}+\sum_{i=1}^{e} \left( \ell_i^{\star,\gamma}\zeta_i  - |\ell_i^{\star,\gamma}| c_{i} \1_{d}  \right) +\sum_{t=0}^T \eta_t^{\star,\gamma}-\xi\right)^-\right]=0,\nonumber
\ee
where $(\ell^{\star,\gamma}$, $\eta^{\star,\gamma})$ is an optimal semi-static strategy to the problem \eqref{eq:util_max_dual_reform1} under the risk aversion level $\gamma$.
\end{Proposition}

\proof
Let us set $\Gamma_{\gamma}:= \pi(\xi)\1_{d}+\sum_{i=1}^{e} \left( \ell_i^{\star,\gamma}\zeta_i  - |\ell_i^{\star,\gamma}| c_{i} \1_{d}  \right) +\sum_{t=0}^T \eta_t^{\star,\gamma}-\xi$ and it follows by \eqref{eq:util_max_dual_reform1} that
\begin{equation}\label{indffrhs}
\sup_{\mathbb{P}\in\mathcal{P}}\log\mathbb{E}^{\mathbb{P}}[e^{-\gamma \Gamma_{\gamma}}]= \sup_{ (\Q,Z) \in \CPSt_e^*} \left\{ \gamma\E^{\Q} \big[ \xi \cdot Z_T\big]-\gamma \pi(\xi)-\Ec(\Q,\Pc)\right\}.
\end{equation}
If $\pi(\xi)=+\infty$, it is clear that $\sup_{\mathbb{P}\in\mathcal{P}}\log\mathbb{E}^{\mathbb{P}}[e^{-\gamma \Gamma_{\gamma}}]=-\infty$. Otherwise, if $\pi(\xi)<+\infty$, it follows by item $\mathrm{(ii)}$ of Lemma \ref{basicpropind} that $\pi_{\gamma}(\xi)$ is increasing in $\gamma$ and moreover $\pi_{\gamma}(\xi)\leq \pi(\xi)$. Therefore, it yields that  $\sup_{\mathbb{P}\in\mathcal{P}}\log\mathbb{E}^{\mathbb{P}}[e^{-\gamma \Gamma_{\gamma}}]\leq 0$ and hence $\mathbb{E}^{\mathbb{P}}[e^{-\gamma\Gamma_{\gamma}}]\leq 1$ uniformly for all $\mathbb{P}\in\mathcal{P}$. By Jensen's inequality, we have
\be
\sup_{\mathbb{P}\in\mathcal{P}}\mathbb{E}^{\mathbb{P}}[\Gamma^-_{\gamma}]\leq \frac{1}{\gamma}\sup_{\mathbb{P}\in\mathcal{P}}\log\mathbb{E}^{\mathbb{P}}[e^{\gamma\Gamma^-_{\gamma}}]
	~\leq~ \frac{1}{\gamma} \sup_{\mathbb{P}\in\mathcal{P}}\log (1+\mathbb{E}^{\mathbb{P}}[e^{-\gamma\Gamma_{\gamma}}]),\nonumber
\ee
which completes the proof.
\qed

\begin{Remark}
Similar results have been obtained in Corollary $5.1$ and Theorem $5.2$ of \cite{Del6authors} in the classical dominated frictionless market model. Thanks to the convex duality \eqref{eq:util_max_dual_reform1}, this paper makes nontrivial extension of the asymptotic convergence on risk aversion level to the setting with both proportional transaction costs and model uncertainty.
\end{Remark}

Again, based on the convex duality representation obtained in the enlarged space, the continuity property and Fatou property of the indifference prices can be shown in the following sense.
\begin{Proposition}
$\mathrm{(i)}$ If $(\xi_n)_{n\in\mathbb{N}}$ is a sequence of option payoffs such that
\be\label{contingconvrg}
\sup_{(\mathbb{Q},Z)\in\mathcal{S}_e^{\ast}}\mathbb{E}^{\mathbb{Q}}[(\xi_n-\xi)\cdot Z_T]\rightarrow 0\ \ \text{and}\ \ \inf_{(\mathbb{Q},Z)\in\mathcal{S}_e^{\ast}}\mathbb{E}^{\mathbb{Q}}[(\xi_n-\xi)\cdot Z_T]\rightarrow 0.
\ee
then $\pi_{\gamma}(\xi_n)\rightarrow \pi_{\gamma}(\xi)$ for any $\gamma>0$.

\vspace{2mm}

\noindent $\mathrm{(ii)}$ For $\xi_n\geq 0$, we have
\be
\pi_{\gamma}(\lim\inf_{n} \xi_n)\leq \lim\inf_n \pi_{\gamma}(\xi_n).
\ee
$\mathrm{(iii)}$ If $(\xi_n)_{n\in\mathbb{N}}$ is a sequence of option payoffs such that $\xi_n {\nearrow} \xi$, $\mathbb{P}$-a.s., then $\pi_{\gamma}(\xi_n){\nearrow} \pi_{\gamma}(\xi)$.

\end{Proposition}

\proof
(i) Recall that $
		\pi_{\gamma}(\xi)=\sup_{ (\Q,Z) \in \CPSt_e^*} \left\{ \E^{\Q} \big[ { \xi \cdot Z_T }\big] -\frac{1}{\gamma} \Ec(\Q,\Pc)\right\}- \sup_{ (\Q,Z) \in \CPSt_e^*} \left\{-\frac{1}{\gamma} \Ec(\Q,\Pc)\right\}$ in \eqref{indifftwoequiv}, we can obtain that
\be
\begin{split}
|\pi_{\gamma}(\xi_n)-\pi_{\gamma}(\xi)|&=\Big|\sup_{ (\Q,Z) \in \CPSt_e^*} \left\{ \E^{\Q} \big[ {  \xi_n \cdot Z_T } \big] -\frac{1}{\gamma} \Ec(\Q,\Pc)\right\}-\sup_{ (\Q,Z) \in \CPSt_e^*} \left\{ \E^{\Q} \big[ { \xi \cdot Z_T } \big] -\frac{1}{\gamma} \Ec(\Q,\Pc)\right\}\Big|\\
&\leq \sup_{ (\Q,Z) \in \CPSt_e^*} |\mathbb{E}^{\mathbb{Q}}[(\xi_n-\xi)\cdot Z_T]|.\nonumber
\end{split}
\ee
The continuity $\pi_{\gamma}(\xi_n)\rightarrow \pi_{\gamma}(\xi)$ follows directly by \eqref{contingconvrg}.\\
\ \\
(ii) The Fatou property can be derived by observing that
\be
\begin{split}
\pi_{\gamma}(\lim\inf_{n} \xi_n)&=\sup_{ (\Q,Z) \in \CPSt_e^*} \left\{ \E^{\Q} \big[ { \lim\inf_n\xi_n \cdot Z_T }\big] -\frac{1}{\gamma} \Ec(\Q,\Pc)\right\}- \sup_{ (\Q,Z) \in \CPSt_e^*} \left\{-\frac{1}{\gamma} \Ec(\Q,\Pc)\right\}\\
&{\leq} \sup_{ (\Q,Z) \in \CPSt_e^*} \left\{ { \lim\inf_n \E^{\Q} \big[ \xi_n \cdot Z_T\big] } -\frac{1}{\gamma} \Ec(\Q,\Pc)\right\}- \sup_{ (\Q,Z) \in \CPSt_e^*} \left\{-\frac{1}{\gamma} \Ec(\Q,\Pc)\right\}\\
&\leq \lim\inf_n \left(\sup_{ (\Q,Z) \in \CPSt_e^*} \left\{ \E^{\Q} \big[ \xi_n \cdot Z_T\big] -\frac{1}{\gamma} \Ec(\Q,\Pc)\right\}- \sup_{ (\Q,Z) \in \CPSt_e^*} \left\{-\frac{1}{\gamma} \Ec(\Q,\Pc)\right\} \right)\\
&=\lim\inf_n \pi_{\gamma}(\xi_n).\nonumber
\end{split}
\ee
(iii) By the Fatou property from part (ii) and item $\mathrm{(vi)}$ of Lemma \ref{basicpropind}, we have
\be
\pi_{\gamma}(\xi)\geq \lim\inf_n\pi_{\gamma}(\xi_n)\geq \pi_{\gamma}(\xi),\nonumber
\ee
which completes the proof.
\qed

\section{Proof of main results}
\label{sec:proof}

This section provides the technical arguments to establish the convex duality \eqref{eq:util_max_dual_reform1} and we shall first work in the fictitious frictionless market on the enlarged space. All three results, namely the convex duality theorem, the dynamic programming principle and the existence of the optimal portfolio will be confirmed. Translating the transaction costs into additional randomness on the enlarged space in both primal and dual problems plays a crucial role to develop some key equivalences.

\subsection{Reformulation of the dual problem}
	As a first step to reduce the complexity of the proof,
	the standard dual problem based on CPS in the model with transaction costs will be reformulated on the enlarged dual space.
        Define
	$$
		\Qcb^* ~:= \big\{ \Qb \in \Qcb_0 ~: \E^{\Qb}\big[ (\xi \cdot X_T)_- \big]+\Ec(\Qb,\Pcb_{\mathrm{int}}) < \infty \big\},
	$$
	where $\Ec(\Qb, \Pcb_{\mathrm{int}})$ is defined exactly as $\Ec(\Q, \Pc)$ in \eqref{eq:def_entropy}.
	For any universally measurable random variable $\varphi: \Omb \to  \R_+$, we further define
	\begin{equation} \label{eq:def_Q_star_phi}
		\Qcb_{\varphi}^* ~:= \big\{ \Qb \in \Qcb^* ~: \E^{\Qb}\big[ \varphi \big] < \infty \big\}
	        ~~\mbox{and}~
		\Qcb_{\varphi}^{*}(0,\theta_{0})
		~:=~
		\big\{ \Qb \in \Qcb_{\varphi}^* ~: \Qb[\Theta_0 = \theta_{0} ] = 1 \big \}.
	\end{equation}
	The function $\varphi$ will be chosen depending on the context,
	it allows to control the integrability of some extra random variables
	when one considers the subsets of $\Qcb^*$ and also in some iteration arguments.

	\begin{Lemma} \label{lemm:equiv_entropy}
		For any universally measurable random vector $\xi: \Om \to \R^d$, one has
		\b*
			 \sup_{ (\Q,Z) \in  \CPSt^*_0} \big\{ \E^{\Q} \big[ \xi \cdot Z_T\big] -\Ec(\Q,\Pc) \big \}
			 ~=~
			 \sup_{\Qb \in \Qcb^*} \big\{ \E^{\Qb} \big[\xi \cdot X_T] - \Ec(\Qb,\Pcb_{\mathrm{int}}) \big \}.
		\e*
	\end{Lemma}
	\proof First, for a given  $(\Q, Z) \in \CPSt^*_0$,
	we associate the probability kernel:
	$$
	q^Z: \om \in \Om \mapsto q^Z(\cdot|\om):=\delta_{(Z/S)(\om)} \in \Bf(\Lambda),
	$$	
	and define $\Qb:=\Q\otimes q^Z$. The construction implies that $\E^{\Q} \big[ \xi \cdot Z_T\big] = \E^{\Qb} \big[\xi \cdot X_T]$ and that   $\Qb \in \Qcb^*$.
	Moreover, for every $\P \in \Pc$, one can similarly define $\Pb:=\P\otimes q^Z\in \overline \Pc_{\mathrm{int}}$.
        If $\Q \ll \P$, one has $\Qb \ll \Pb$ and $d\Q/d\P=d \Qb/d \Pb$, $\Pb$-a.s.
	If $\Q \ll \P$ is not true, then $\Ec(\Q, \P) = \infty$ by definition.
	This implies that $\Ec(\Q, \Pc) \ge \Ec(\Qb, \Pcb_{\mathrm{int}})$.
	Therefore,
	\b*
		 \sup_{ (\Q,Z) \in \CPSt^*_0} \big\{ \E^{\Q} \big[ \xi \cdot Z_T\big] -\Ec(\Q,\Pc) \big \}
		 ~\le~
		\sup_{\Qb \in \Qcb^*} \big\{ \E^{\Qb} \big[\xi \cdot X_T] - \Ec(\Qb,\Pcb_{\mathrm{int}}) \big \}.
	\e*
	Conversely, let us fix $\Qb \in \Qcb^*$, and define $\Q := \Qb|_{\Om}$ and $Z_t:= \E^{\Qb} \big[ X_{t} \big| \Fc_t \big]$ for $t \leq T$. As $\Qb \ll \Pb$ for some $\Pb \in \Pcb_{\mathrm{int}}$, then $\Q \ll \P :=  \Pb|_{\Om}\in \Pc$.
	Moreover, the fact that $X$ is an $(\Fbb, \Qb)$-martingale implies that $Z$ is an $(\F,\Q)$-martingale.
	Then, $(\Q, Z) \in \CPSt^*_0$ and  $\E^{\Q} \big[ \xi \cdot Z_T\big] = \E^{\Qb} \big[\xi \cdot X_T]$.
	Now as $d\Q/d\P=\E^{\Pb}[d\Qb/d\Pb| \Fc_{T}]$ and $x \mapsto x \log(x)$ is convex on $\R_+$, we have $\Ec(\Q, \P) \le \Ec(\Qb, \Pb)$ by Jensen's inequality.
	It follows that
		\b*
		 \sup_{ (\Q,Z) \in  \CPSt^*_0} \big\{ \E^{\Q} \big[ \xi \cdot Z_T\big] -\Ec(\Q,\Pc) \big \}
		 ~\ge~
		\sup_{\Qb \in \Qcb^*} \big\{ \E^{\Qb} \big[\xi \cdot X_T] - \Ec(\Qb,\Pcb_{\mathrm{int}}) \big \},
	\e*
	and we hence conclude the proof.
	 \qed

\subsection{Proof of Theorem \ref{thm:utility_max_duality}(Case $e=0$)}
\label{subsec:proof_e0}

	In view of Lemma \ref{lemm:equiv_entropy} and Proposition \ref{prop:reformulation},
	one can first establish the duality result of the utility maximization problem on the enlarged space $\Omb$,
	in order to prove Theorem \ref{thm:utility_max_duality}.

	\begin{Proposition} \label{prop:enlarged_utility_max_duality}
                 Let $g:=\xi \cdot X_T$ and  $\NA(\Pcb_{\mathrm{int}})$ hold true.
		Then for any universally measurable random variable $\varphi: \Omb \to \R_+$, one has
		\be  \label{eq:max_util_duality_enlarg}
			\overline V
			~:=~
			\inf_{H \in \Hc} \sup_{\Pb \in \Pcb_{\mathrm{int}}} \log \E^{\Pb} \big[ \exp\big( g + (H \circ X)_T \big) \big]
			\!\!&=&\!\!
			\sup_{\Qb \in \Qcb^*} \big\{ \E^{\Qb} \big[ g \big] - \Ec(\Qb,\Pcb_{\mathrm{int}}) \big \}\\
			\!\!&=&\!\!
			 \sup_{\Qb \in \Qcb^*_{\varphi}} \big\{ \E^{\Qb} \big[ g \big] - \Ec(\Qb,\Pcb_{\mathrm{int}}) \big \}. \nonumber
		\ee
		Moreover, the infimum of the problem $\overline V$ is attained by some optimal trading strategy $\widehat H \in \Hc$.
	\end{Proposition}
	
	\begin{Remark} \label{Rmk:improve}
	The above duality result is similar to that in \cite{bartl2016exponential},
	but differs substantially with theirs in the following two points:
	
	\vspace{2mm}
	
	\noindent \rmi In our current work, we have relaxed the strong one-period no-arbitrage condition for all $\om_t \in \Om_t$ assumed in \cite{bartl2016exponential}.
	Indeed, the strong no-arbitrage condition is needed in \cite{bartl2016exponential} because their duality and dynamic programming are mixed with each other.
	 More precisely, with the notations in \cite[Section 4]{bartl2016exponential},
	they need the relation
	``$\Ec_t(\om,x) = \Dc_t(\om) + x$'' to hold
	for all $t$ and $\om \in \Om_t$ to guarantee the measurability of $\Ec_t$ through $\Dc_t$
	(see in particular their equation (21) and their Proof of Lemma 4.6). In Appendix \ref{subsec:nondominatedUtility}, we shall give more details on this point.

	\vspace{2mm}

	\noindent \rmii It is worth noting that the reformulations in Proposition \ref{prop:reformulation} on the enlarged space do not exactly correspond to  standard quasi-sure utility maximization problem.
		Indeed, we still restrict the class of strategies to $\Ft$-predictable processes, as opposed to $\Fbb$-predictable processes.
		The fact that the formulation with these two different filtrations are equivalent will be proved by using a minimax argument.
	\end{Remark}

	\noindent {\bf Proof of Theorem \ref{thm:utility_max_duality} (case $e=0$)}
		First, using Lemma \ref{lemm:equiv_entropy} and Proposition \ref{prop:reformulation},
		the duality \eqref{eq:util_max_dual_reform1} can be deduced immediately from \eqref{eq:max_util_duality_enlarg} in Proposition \ref{prop:enlarged_utility_max_duality}.
		Moreover, given the optimal trading strategy $\widehat H \in \Hc$ in Proposition \ref{prop:enlarged_utility_max_duality}, we can construct $\hat \eta$ by \eqref{eq:H2eta} and
		show its optimality by almost the same arguments as in Step 2 of Proposition \ref{prop:reformulation} $\mathrm{(ii)}$.
	\qed

	\vspace{0.5em}
	
	In the rest of Section \ref{subsec:proof_e0}, we will provide the proof of Proposition \ref{prop:enlarged_utility_max_duality} in several steps.

\paragraph{The weak duality} As in the classical results, one can easily obtain a weak duality result.

	\begin{Lemma} \label{lemm:weak_dual_utility}
		For any universally measurable function $g: \Omb \to \R \cup \{\infty\}$, one has
		$$
			\inf_{H \in \Hc} \sup_{\Pb \in \Pcb_{\mathrm{int}}} \log \E^{\Pb} \big[ \exp\big( g + (H \circ X)_T \big) \big]
			~\ge~
			\sup_{\Qb \in \Qcb^*} \big\{ \E^{\Qb} [g] - \Ec(\Qb,\Pcb_{\mathrm{int}}) \big \}.
		$$
	\end{Lemma}
	\proof Using the result in the \cite[Proof of Theorem 4.1 - dynamic programming principle]{bartl2016exponential},
	one knows that for any $H \in \Hc$, $\Pb \in \Pcb_{\mathrm{int}}$ and $\Qb \in \Qcb^*$, one has
	$$
		\log \E^{\Pb} \big[ \exp\big( g + (H \circ X)_T \big) \big]
		~\ge~
		 \E^{\Qb} [g] - \Ec(\Qb,\Pb).
	$$
	(Note that $\Ec(\Qb, \Pb) = \infty$ if $\Qb$ is not dominated by $\Pb$.)
	Therefore it is enough to take supremum over $\Qb$ (and $\Pb$) and then take infimum over $H \in \Hc$ to obtain the two weak duality results in the claim.
	\qed

	\vspace{2mm}

	We can next turn to (and for the duality, it suffices to) prove that
	\be \label{eq:max_util_duality_enlarg_p}
		 \inf_{H \in \Hc} \sup_{\Pb \in \Pcb_{\mathrm{int}}} \log \E^{\Pb} \big[ \exp\big( g + (H \circ X)_T \big) \big]
		~\le~
		\sup_{\Qb \in \Qcb^*_\varphi} \big\{ \E^{\Qb} \big[g] - \Ec(\Qb,\Pcb_{\mathrm{int}}) \big \},
	\ee
	for any universally measurable random variable $\varphi: \Omb \to [0, \infty)$.

\paragraph{The one-period case $T=1$}
	Let us first consider the one-period case $T=1$.
	Define
		$$
			 \Lambda_{\rm int}(0, \om_{0}) := \{\theta_{0} \in \Lambda_{1}: S_{0}(\om_0) \theta_{0}\in {\rm int} K^{*}_{0} \},
		$$
		and for each $\theta_{0} \in  \Lambda_{\rm int}(0, \om_{0})$,
		$$
			\Pcb_{\mathrm{int}}^{\delta}(0,\theta_{0})
			\!:=\!
			\big\{ \Pb \in \Pcb_{\mathrm{int}} ~: \Pb[ \Theta_0 = \theta_{0} ] = 1 \big\}.
		$$
		 Define $\NA(\Pcb_{\mathrm{int}}^{\delta}(0,\theta_{0}))$ as $\NA(\Pcb_{\mathrm{int}})$  in Definition \ref{def:NA}  with $\Pcb_{\mathrm{int}}^{\delta}(0,\theta_{0})$ in place of  $\Pcb_{\mathrm{int}}$.
		Then,
		$\NA(\Pcb_{\mathrm{int}})$ implies that $\NA(\Pcb_{\mathrm{int}}^{\delta}(0,\theta_{0}))$ holds for  every
		$\theta_{0}\in \Lambda_{\rm int}(0, \om_{0})$.


	\begin{Lemma} \label{lemm:utili_max_dual_T1}
		Let $T = 1$, and $g_1: \Omb \to \R \cup \{\infty\}$ be upper semi-analytic and also $(\om, \theta_0, \theta_1)\in \Om\x \Lambda_{1}\x \Lambda_{1}\to g_1(\om, \theta_0, \theta_1)$ depend only on $(\om, \theta_1)$.
		Assume that $\NA(\Pcb_{\mathrm{int}})$ holds.
		Then, for $g=g_{1}$, the inequality \eqref{eq:max_util_duality_enlarg_p} holds for any universally measurable random variable $\varphi: \Omb \to [0, \infty)$ and both terms are not equal to $-\infty$.
		Moreover, there exists an optimal solution $\widehat H\in \Hc$ for the infimum problem at the left hand side.
		{ In consequence, Proposition \ref{prop:enlarged_utility_max_duality} holds true for the case $T=1$.}
	\end{Lemma}
	\proof \textit{Step 1}: Although the context is slightly different, we can still follow the same arguments line by line in step (b) of the proof of Theorem 3.1 and Lemma 3.2 of \cite{bartl2016exponential} to obtain the existence of the optimal strategy $\widehat H$
	(see also the proof of Theorem 2.2 of \cite{nutz2016utility}), where the key argument is to show that $h \mapsto \sup_{\Pb \in \Pcb_{\mathrm{int}}} \log \E^{\Pb} \big[ \exp( g+ h (X_1 - X_0)) \big]$ is lower-semicontinuous.
	
	\vspace{1mm}
	
	\noindent \textit{Step 2}: We then turn to prove the duality result. First, notice that $\Hc = \R^d$ when $T=1$,
	 and that $(g_1, X_1)(\om, \theta_0, \theta_1)$ is independent of $\theta_0$.
	Then, for all $\theta_{0} \in \Lambda_{\rm int}(0, \om_{0})$,
	\be \label{eq:Pcb_ind_theta}
		\big\{ \Pb \circ (g_1, X_1)^{-1} : \Pb \in \Pcb_{\mathrm{int}}(0,\theta_{0}) \big\}
		~=~
		\big\{ \Pb \circ (g_1, X_1)^{-1} : \Pb \in \Pcb_{\mathrm{int}}(0,\1) \big\},
	\ee
	where $\1$ represents the vector of $\R^{d}$ with all entries equal to $1$.
	Thanks to the standard concatenation argument, it is clear that
	$$
		\overline V ~=~  \inf_{h_1 \in \R^d} \sup_{\theta_{0}\in \Lambda_{\rm int}(0, \om_{0})} \sup_{\Pb \lll \Pcb_{\mathrm{int}}^{\delta}(0,\theta_{0})} \log \E^{\Pb} \big[ \exp(g_1 + h_1 \cdot X_1 - h_1 \cdot S_0 \theta_0) \big].
	$$

	To proceed, let us first assume that $g$ is bounded from above, while the general case will be treated later. 
	We define the function
	$$
		\alpha(h_1, \theta_0)
		~:=~
		\sup_{\Pb \lll \Pcb_{\mathrm{int}}^{\delta}(0,\theta_{0})} \!\!\!
		\log \E^{\Pb} \big[ \exp ( g_1+ h_1 \cdot X_1 - h_1 \cdot S_{0}\theta_{0}) \big].
	$$
	It is clear that $\alpha(h_1, \theta_0) > -\infty$.
	We then define its effective domain (which is independent of $\theta_0$ by \eqref{eq:Pcb_ind_theta})	
	$$
		D ~:=~
		\Big\{ h \in \R^d ~: 
			\alpha(h, \theta_0) < \infty
		\Big\}.
	$$
	Observe that $0 \in D$, which implies that $D \neq \emptyset$.
	Next, by H\"older inequality, one has, for all $\lambda \in (0,1)$ and (universally measurable) random variables $Y_1$ and $Y_2$,
	$$
		\E^{\Pb} \big[ \exp \big( \lambda Y_1+ (1-\lambda)Y_2 \big) \big]
		~\leq~
		\Big( \E^{\Pb} \big[\exp(Y_1) \big] \Big)^{\lambda}\Big( \E^{\Pb} \big[\exp(Y_2) \big] \Big)^{1-\lambda},
	$$
	and hence
	\be \label{eq:conv_log_exp}
		\log \E^{\Pb} \big[ \exp \big( \lambda Y_1+ (1-\lambda) Y_2 \big) \big]
		~\leq~
		\lambda  \log \E^{\Pb} \big[\exp(Y_1) \big] + (1- \lambda) \log \E^{\Pb} \big[\exp(Y_2) \big] .
	\ee
	Then for any $h_1, h_2 \in D$ and $h :=  \lambda h_1 + (1-\lambda) h_2$, one has for any $\Pb  \lll \Pcb_{\mathrm{int}}^{\delta}(0,\theta_{0})$,
	\b*
		&&
		\log \E^{\Pb} \big[ \exp ( g_1+ h \cdot X_1 - h \cdot S_{0}\theta_{0}) \big] \\
		&\le&
			\lambda
			\log \E^{\Pb} \big[ \exp ( g_1+ h_1 \cdot X_1 - h_1 \cdot S_{0}\theta_{0}) \big]
			+
			(1-\lambda)
			\log \E^{\Pb} \big[ \exp ( g_1+ h_2 \cdot X_1 - h_2 \cdot S_{0}\theta_{0}) \big]
		\\
		&\le&
		\lambda \alpha(h_1, \theta_0) + (1-\lambda) \alpha(h_2, \theta_0) < \infty.
	\e*
	This implies that $h \in D$ and hence $D$ is a convex subset in $\R^d$.
	
	We further notice that, for all $h_1 \in D$,
	$$
		\theta_{0}
		~\mapsto~
		\alpha(h_1, \theta_0)
		=
		\sup_{\Pb \lll \Pcb_{\mathrm{int}}^{\delta}(0,\1)} \!\!\! \log \E^{\Pb} [ \exp ( g_1+ h_1 \cdot X_1 - h_1 \cdot S_{0}\theta_{0})]
		~\mbox{is affine}.
	$$
	and $h_1 \in D \mapsto \alpha(h_1, \theta_0) \in \R$ is convex in view of inequality \eqref{eq:conv_log_exp} again.
	We can then use the minimax theorem to deduce that
	\b*
		\overline V
		&=&
		 \inf_{h_1 \in \R^d} \sup_{\theta_{0}\in \Lambda_{\rm int}(0, \om_{0})} \alpha(h_1, \theta_0)
		=
		 \inf_{h_1 \in \R^d} \sup_{\theta_{0}\in \Lambda(0,\om_{0})} \alpha(h_1, \theta_0)
		\\
		&=&
		 \inf_{h_1 \in D} \sup_{\theta_{0}\in \Lambda(0,\om_{0})} \alpha(h_1, \theta_0)
		=
		 \sup_{\theta_{0}\in \Lambda(0,\om_{0})}  \inf_{h_1 \in D} \alpha(h_1, \theta_0)					 
		=
		 \sup_{\theta_{0}\in \Lambda(0,\om_{0})}  \inf_{h_1 \in \R^d} \alpha(h_1, \theta_0)		
		\\
		&=&
		\sup_{\theta_{0}\in \Lambda_{\rm int}(0, \om_0)} \inf_{h_1 \in \R^d} \sup_{\Pb \lll \Pcb_{\mathrm{int}}^{\delta}(0,\theta_{0})} \log \E^{\Pb} \big[ \exp(g_1 + h_1 \cdot X_1 - h_1 \cdot S_0 \theta_0) \big].
	\e*
	In { the above argument},  $\Lambda(0,\om_{0})$ denotes the closure of $\Lambda_{\rm int}(0, \om_{0})$,
	and we can replace $\Lambda_{\rm int}(0, \om_{0})$ by $\Lambda(0,\om_{0})$ since $\theta_0 \mapsto \alpha(h_1, \theta_0)$ is affine,
	and $\theta_0 \mapsto \inf_{h_1 \in \R^d} \alpha(h_1, \theta_0)$ is concave and hence lower semicontinuous(as $\inf_{h_1 \in \R^d} \alpha(h_1, \theta_0)$ cannot take the value $-\infty$ by the weak duality and the definition of $\Qcb^*_\varphi(0,\theta_{0})$ in \eqref{eq:def_Q_star_phi}).
	 Using the one period duality result in \cite[Theorem 3.1]{bartl2016exponential}, we obtain
	$$
		\overline V = \sup_{\theta_{0}\in \Lambda_{\rm int}(0, \om_{0})} \sup_{\Qb \in \Qcb_{\varphi}^{*}(0,\theta_{0})}
		\Big \{ \E^{\Qb} [g_1] - \Ec\big( \Qb, \Pcb_{\mathrm{int}}^{\delta}(0,\theta_{0}) \big) \Big\}.
	$$

	For the case that $g$ is not necessarily bounded from above, one notices that both
	$$
		g_1 \mapsto  \inf_{h_1 \in \R^d} \sup_{\theta_{0}\in \Lambda_{\rm int}(0, \om_{0})} \sup_{\Pb \lll \Pcb_{\mathrm{int}}^{\delta}(0,\theta_{0})} \log \E^{\Pb} \big[ \exp(g_1 + h_1 \cdot X_1 - h_1 \cdot S_0 \theta_0) \big]
	$$
	and
	$$
		g_1 \mapsto  \inf_{h_1 \in \R^d} \sup_{\Pb \lll \Pcb_{\mathrm{int}}^{\delta}(0,\theta_{0})} \log \E^{\Pb} \big[ \exp(g_1 + h_1 \cdot X_1 - h_1 \cdot S_0 \theta_0) \big]
	$$
	are continuous from below by Step (b) of Lemma 3.2. of \cite{bartl2016exponential}.
	Define, for any $n \ge 1$,
	$$
		\alpha_n(h_1, \theta_0)
		~:=~
		\sup_{\Pb \lll \Pcb_{\mathrm{int}}^{\delta}(0,\theta_{0})} \!\!\!
		\log \E^{\Pb} \big[ \exp (g_1 \wedge n + h_1 \cdot X_1 - h_1 \cdot S_{0}\theta_{0}) \big].
	$$
	It follows that 
	\b*
		\overline V 
		&=& 
		\sup_{n \ge 1}  \inf_{h_1 \in \R^d} \sup_{\theta_{0}\in \Lambda_{\rm int}(0, \om_{0})} \alpha_n(h_1, \theta_0)
		~=~
		\sup_{n \ge 1} 
		\sup_{\theta_{0}\in \Lambda_{\rm int}(0, \om_0)} \inf_{h_1 \in \R^d} \alpha_n(h_1, \theta_0)\\
		&=&
		\sup_{\theta_{0}\in \Lambda_{\rm int}(0, \om_0)}
		\sup_{n \ge 1} 
		\inf_{h_1 \in \R^d} \alpha_n(h_1, \theta_0)
		~=~
		\sup_{\theta_{0}\in \Lambda_{\rm int}(0, \om_0)}
		\inf_{h_1 \in \R^d} \alpha(h_1, \theta_0)\\
		&=&
		\sup_{\theta_{0}\in \Lambda_{\rm int}(0, \om_0)} \inf_{h_1 \in \R^d} \sup_{\Pb \lll \Pcb_{\mathrm{int}}^{\delta}(0,\theta_{0})} \log \E^{\Pb} \big[ \exp(g_1 + h_1 \cdot X_1 - h_1 \cdot S_0 \theta_0) \big],
	\e*
	which completes the proof of this step.
	
	\vspace{0.5em}
	
	\noindent \textit{Step 3}: To conclude the proof, it is enough to prove that
	\be \label{eq:conditioning_entropy}
		\sup_{\theta_{0}\in \Lambda_{\rm int}(0, \om_{0})} \sup_{\Qb \in \Qcb_{\varphi}^{*}(0,\theta_{0})}
		\Big \{ \E^{\Qb} [g_1] - \Ec\big( \Qb, \Pcb_{\mathrm{int}}^{\delta}(0,\theta_{0}) \big) \Big\}
		~\ge~
		\sup_{\Qb \in \Qcb_{\varphi}^*} \big\{ \E^{\Qb} \big[g_1] - \Ec(\Qb,\Pcb_{\mathrm{int}}) \big \},
	\ee
	as the reverse inequality is trivial by the fact that  $\Qcb_{\varphi}^{*}(0,\theta_{0}) \subset \Qcb_{\varphi}^*$
	and that $\Ec\big( \Qb, \Pcb_{\mathrm{int}}^{\delta}(0,\theta_{0}) \big) =  \Ec(\Qb,\Pcb_{\mathrm{int}})$ for all $\Qb \in \Qcb_{\varphi}^{*}(0,\theta_{0})$.
	Let $\Qb \in \Qcb_{\varphi}^*$ and denote by $(\Qb_{\theta_0})_{\theta_0 \in \Lambda_{\rm int}(0, \om_{0})}$ a family of r.c.p.d. of $\Qb$ knowing $\theta_0$, then by  \cite[Lemma 4.4]{bartl2016exponential}, we have
	\b*
		\E^{\Qb} \big[g_1] - \Ec(\Qb,\Pcb_{\mathrm{int}})
		&=&
		\E^{\Qb} \Big[
			\E^{\Qb_{\theta_0}} \big[ g_1 \big]
			- \Ec \big( \Qb_{\theta_0}, \Pcb_{\mathrm{int}}^{\delta}(0,\theta_{0}) \big)
		\Big]
		-
		\Ec \big( \Qb \circ \theta_0^{-1}, \Pcb_{\mathrm{int}}|_{\Omb_0}  \big) \\
		&\le&
		\sup_{\theta_{0}\in \Lambda_{\rm int}(0, \om_{0})} \sup_{\Qb \in \Qcb_{\varphi}^{*}(0,\theta_{0})}
		\Big \{ \E^{\Qb} [g_1] - \Ec\big( \Qb, \Pcb_{\mathrm{int}}^{\delta}(0,\theta_{0})  \big) \Big\}.
	\e*
	Taking the supremum over $\Qb$ in $\Qcb^*_{\varphi}$, we verify \eqref{eq:conditioning_entropy}.
	\qed

\paragraph{The multi-period case: measurable selection of the dynamic strategy}
	Let us extend the above definitions of  $\Lambda_{\rm int}(0, \om_{0})$, $\Pcb_{\mathrm{int}}^{\delta}(0,\theta_{0})$ and $\Qcb_{\varphi}^{*}(0,\theta_{0})$
	to an arbitrary initial time $t$ and initial path $\omb^{t}$.
	For $t \ge 1$ and $\omb = \omb^t = (\om^t, \theta^t) \in \Omb_t$, let us first recall the definition of $\Lambda_{\rm int} (t, \om^{t}) $ :
	$$
		\Lambda_{\rm int} (t, \om^{t})
		~:=~
		\{\theta_t \in \Lambda_1 ~: S_t(\om^t) \theta_t \in \mathrm{int}K^*_t(\om^t) \}
		~\subset~
		\Lambda_1.
	$$
	Note that $\Pcb_{\mathrm{int}}(t,\omb) \subset \Bf(\Om_1 \x \Lambda_1)$ is defined in \eqref{eq:def_Pcb_t}.
	We introduce
	\be \label{eq:Pcb_delta_int}
		\Pcb^{\delta}_{\mathrm{int}}(t,\omb)  := \big\{
		\delta_{\omb^{t}} \otimes \Pb_{t+1} ~:  \Pb_{t+1} \in \Pcb_{\mathrm{int}}(t,\omb) \big\},
	\ee
	and
	$$
		\Pct^{\delta}_{\mathrm{int}}(t,\om)
		:=
		\left\{
			(\delta_{\om^{t}}\times \mu(d \theta^t)) \otimes \Pb_{t+1}
			~:  \Pb_{t+1} \in \Pcb_{\mathrm{int}}(t,\omb), ~\mu
				\in \Bf \big(\Lambda_{\mathrm{int}}(0,\om_0) \x \cdots \x \Lambda_{\mathrm{int}}(t,\om^t) \big)
		\right\},
	$$
	where the latter consists in a version of $\Pcb^{\delta}_{\mathrm{int}}(t,\omb)$ in which $\theta^{t}$ is not fixed anymore.
		
	\begin{Remark}\label{rem: NA Pintdelta omega}	
		\noindent {\rm (i)} For a fixed $\om \in \Om_t$, let us define $\NA(\Pct^{\delta}_{\mathrm{int}}(t,\om))$ by
		$$
			h(X_t) \cdot(X_{t+1}-X_{t}) \ge 0,~ \Pct^{\delta}_{\mathrm{int}}(t,\om)\mbox{-q.s.}
			~\Longrightarrow~
			h(X_t) \cdot(X_{t+1}-X_{t})= 0,~ \Pct^{\delta}_{\mathrm{int}}(t,\om)\mbox{-q.s.},
		$$
		for every universally measurable function  $h: \R^{d} \to \R^d$.
		By applying Proposition \ref{prop:equiv_NA} with $\Pc(t,\om)$ in place of $\Pc$, one obtains that
		$\NA2(t, \om)$ defined in \eqref{eq:NA2t} is equivalent to $\NA(\Pct^{\delta}_{\mathrm{int}}(t,\om))$.
		
		\vspace{1mm}

		\noindent {\rm (ii)}
	We recall that for each $t \le T$ and $\om \in \Om_t$, the condition $\NA2(t, \om)$ is satisfied if
	\be \label{eq:NA2t}
		\zeta \in K_{t+1}(\om, \cdot),~ \Pc_t(\om) \mbox{-q.s.}
		~~~\mbox{implies}~~
		\zeta \in K_t(\om), ~~\mbox{for all}~\zeta \in \R^d.
	\ee
	Then by \cite[Lemma 3.6]{bouchard2016consistent},
		the set $N_t := \{\om : \NA2(t,\om) ~\mbox{fails}\}$ is universally measurable.
		Moreover,  $N_t$ is a $\Pc$-polar set if $\NA2(\Pc)$ holds.

		\vspace{1mm}
		
		\noindent {\rm (iii)}
		It follows from {\rm (i)} and \rm (ii)
		that
		$\NA2(t,\om)$ or equivalently $\Pct^{\delta}_{\mathrm{int}}(t,\om)$ holds for all $\om$ outside a $\Pc$-polar set $N$, whenever $\NA2(\Pc)$ holds.
		The latter is equivalent to $\NA(\Pcb_{\mathrm{int}})$ by  Proposition \ref{prop:equiv_NA}.
		Therefore, if $\NA(\Pcb_{\mathrm{int}})$ holds,
		there exists a $\Pcb_{\mathrm{int}}$-polar $\overline N := N \x \Lambda$, such that for all $\omb = (\om, \theta) \notin \overline N$,
		$\NA(\Pct^{\delta}_{\mathrm{int}}(t,\om))$ holds.

		\vspace{1mm}

		\noindent {\rm (iv)}
		 Finally, for a fixed $\omb \in \Omb_t$, we define $\NA(\Pcb^{\delta}_{\mathrm{int}}(t,\omb))$ by
		$$
			h\cdot(X_{t+1}-X_{t}) \ge 0,~ \Pcb^{\delta}_{\mathrm{int}}(t,\omb)\mbox{-q.s.}
			~~~\Longrightarrow~~~
			h\cdot(X_{t+1}-X_{t})= 0,~ \Pcb^{\delta}_{\mathrm{int}}(t,\omb)\mbox{-q.s.},
		$$
		for every  $h\in  \R^{d}$.
		Then, $\NA(\Pct^{\delta}_{\mathrm{int}}(t,\om))$ implies $\NA(\Pcb^{\delta}_{\mathrm{int}}(t,\om,\theta))$ for all $\theta\in \Lambda$  (see also Remark 3.9 of \cite{BDT17}).
		
	\end{Remark}


	\vspace{0.5em}

	Let us fix a functional $g_{t+1}: \Omb_{t+1} \to \R \cup \{ \infty\}$ which is upper semi-analytic and
	such that
	$g_{t+1}( \om^{t+1}, \theta_0, \cdots, \theta_{t+1})$ depends only on $(\om^{t+1}, \theta_{t+1})$.
	Then for any universally measurable random variable $Y_{t+1}: \Omb_{t+1} \to  \R_+$,
	we introduce
	$$
		\Qcb_{Y_{t+1}}^{*}(t,\omb)
		~:=~\left\{ \Qb \in \Bf(\Omb_{t+1})  ~:~~
		\begin{aligned}
			& \Qb \lll \Pcb^{\delta}_{\mathrm{int}}(t,\omb),~\E^{\Qb}[ X_{t+1} - X_t ] = 0,
			~~\E^{\Qb}[ Y_{t+1} ] < \infty, & \\
			&\E^{\Qb}\big[ g_{t+1}^- + |X_{t+1}-X_t| \big]+\Ec \big(\Qb, \Pcb^{\delta}_{\mathrm{int}}(t,\omb) \big) < \infty &
		\end{aligned}
		\right\},
	$$
	and by setting $Y_{t+1} \equiv 0$, we define
	\be \label{eq:def_g_n}
		g_t(\omb^t)
		~:=
		\sup_{\Qb \in \Qcb_{0}^{*}(t,\omb^t)}
		\Big\{
			\E^{\Qb}[ g_{t+1}] - \Ec \big(\Qb, \Pcb^{\delta}_{\mathrm{int}}(t,\omb^t) \big)
		\Big\}, ~~\mbox{for all}~ \omb^t \in \Omb_t.
	\ee

	\begin{Remark} \label{rem:g_prime}
		Let $\bar \om=(\om,\theta)$ and $\bar \om'=(\om',\theta')$ be such that  $\om^{t} = (\om')^{t}$ and $\theta_t = \theta^{'}_t$.
		Then, it follows from the  definition of $\Pcb^{\delta}_{\mathrm{int}}(t,\omb)$ and $\Qcb_{Y_{t+1}}^{*}(t,\omb)$  that
		$$
			\left\{ \Qb \circ (g_{t+1}, X_t, X_{t+1})^{-1} ~:\Qb \in \Qcb_{Y_{t+1}}^{*}(t,\omb) \right\}
			=
			\left\{ \Qb \circ (g_{t+1}, X_t, X_{t+1})^{-1} ~:\Qb \in \Qcb_{Y_{t+1}}^{*}(t,\omb ') \right\}.
		$$
		Since $g_{t+1}( \om^{t+1}, \theta_0, \cdots, \theta_{t+1})$ is assumed to be independent of $(\theta_0, \cdots, \theta_t)$,
		then it is clear that
		$g_t(\omb^{t})$ depends only on $(\om^{t}, \theta_t)$ for $\omb^{t} = (\om^{t}, \theta_0, \cdots, \theta_t)$.
	\end{Remark}
	
	The above remark allows us to define
	\be \label{eq:def_g_t_p}
		g'_t(\om^t, h)
		~:=
		\sup_{\theta_t \in \Lambda_{\rm int}(t,\om^{t})}
		\big\{ g_t(\om^t, \theta_t) + h \cdot S_t(\om) \theta_{t} \big\},
		~~\forall~(\om^t, h) \in \Om_t \x \R^{d}.
	\ee

	\begin{Remark} \label{rmk:independentY}
	 From Remark \ref{rem: NA Pintdelta omega}, $\NA(\Pcb_{\mathrm{int}})$ implies that $\NA(\Pcb^{\delta}_{\mathrm{int}}(t,\omb))$ holds for $\Pb$-a.e. $\omb \in \Omb$ under any $\Pb \in \Pcb_{\mathrm{int}}$.
		We can in fact apply Theorem 3.1 of \cite{bartl2016exponential} to obtain that
		$$
			g_t(\omb)
			~=~
			\sup_{\Qb \in \Qcb_{Y_{t+1}}^{*}(t,\omb)}
			\Big\{
			\E^{\Qb}[ g_{t+1}] - \Ec \big(\Qb, \Pcb^{\delta}_{\mathrm{int}}(t,\omb) \big)
			\Big\},~
			\Pcb_{\mathrm{int}}\mbox{-q.s.},
		$$
		for all universally measurable random variables $Y_{t+1}: \Omb_{t+1} \to \R_+$.		
	\end{Remark}

	\begin{Lemma} \label{qanalytic}
		For every $t$, the graph set
		\b*
			\Big[ \Big[ \Qcb_{0}^{*}(t) \Big] \Big]
			:=
			\Big\{ (\omb, \Qb): \omb \in \Omb_t,
				\Qb \in \Qcb_{0}^{*}(t,\omb)
			\Big\}
			~\text{is analytic.}
		\e*
	\end{Lemma}
\proof We follow the arguments in Lemma 4.5 of \cite{bartl2016exponential} and Lemma 4.8 of \cite{BouchardNutz.13}.
First, { as} $g_{t+1}\wedge0 +|X_{t+1} - X_t|$ is upper semi-analytic, an application of Proposition 7.48 of \cite{BertsekasShreve.78} shows that $(\omb, \Qb) \mapsto \E^{\Qb}\big[ g_{t+1}\wedge0 -|X_{t+1} - X_t| \big]$ is upper semi-analytic.

 Furthermore, from the definition of $\Pcb^{\delta}_{\mathrm{int}}(t)$ in \eqref{eq:Pcb_delta_int},
	one observes that the graph set $\big[ \big[ \Pcb^{\delta}_{\mathrm{int}}(t) \big] \big]$ is analytic,
	as $\big[ \big[ \Pcb_{\mathrm{int}}(t) \big] \big]$ is analytic (see Remark \ref{remark:panalytic}).
	Then using the Borel measurability of the relative entropy (Lemma 4.2 of \cite{bartl2016exponential}),
	one obtains from a measurable selection argument (see e.g. Proposition 7.47 of \cite{BertsekasShreve.78}) that $(\omb, \Qb) \mapsto -\Ec \big(\Qb, \Pcb^{\delta}_{\mathrm{int}}(t,\omb) \big)$ is upper semi-analytic.

	It follows that
$$
A:= \left\{ (\omb, \Qb): \E^{\Qb}\left[ g_{t+1}\wedge0 - |X_{t+1}-X_t| \right]-\Ec \left(\Qb, \Pcb^{\delta}_{\mathrm{int}}(t,\omb) \right) > - \infty \right\}
$$
is an analytic set.
	{By Lemma 4.8 of \cite{BouchardNutz.13}, we know}
	$$
B(\omb):= \{ (\Qb, \Pb) \in \Bf(\Omb_{t+1}) \times \Bf(\Omb_{t+1}) : \Pb \in \Pcb^{\delta}_{\mathrm{int}}(t,\omb),~ \E^{\Qb}[X_{t+1}-X_t]=0,~ \Qb \ll \Pb \}
$$
has an analytic graph. Notice that the set
$$
C:=  \big\{ \big(\omb, \Qb \big): \Qb \lll \Pcb^{\delta}_{\mathrm{int}}(t,\omb), ~ \E^{\Qb}[X_{t+1}-X_t]=0 \big\}
$$
is the image of the graph set $\big[\big[ B \big]\big]$ under canonical projection $\Omb_t \times \Bf(\Omb_{t+1}) \times \Bf(\Omb_{t+1}) \mapsto \Omb_t \times \Bf(\Omb_{t+1})$ and thus analytic. Finally, it is shown that
$$
\left[ \left[ \Qcb_0^{*}(t) \right] \right] = A \cap C
$$
is analytic.
\qed

	\begin{Lemma} \label{lemm:utility_max_selec_h}
		Assume that  $\NA(\Pcb_{\mathrm{int}})$ holds true.
		 Then both $g_t$ and $g'_t$ are upper semi-analytic, and there is a  universally measurable map $h_{t+1}: \Om_t \x \R^d \to \R^d$ together with a $\Pc$-polar set $N$ such that,
		for every $(\om, h_t) \in N^{c} \x \R^d$, one has
		\b*
			~~g'_t(\om^t, h_t)
			=\!\!\!
			\sup_{\theta_t \in \Lambda_{\rm int}(t,\om^{t})} \sup_{\Pb \in \Pcb^{\delta}_{\mathrm{int}}(t,\omb)}
			\!\!\! \log \E^{\Pb}
			\Big[
				\exp\big( g_{t+1} + h_{t+1}(\om^t, h_t) (X_{t+1} - X_t) + h_t X_t\big)
			\Big]
                         >
                         - \infty.
		\e*
		
	\end{Lemma}
	\proof
	The proof follows the track of measurable selection arguments as in Lemma 3.7 of \cite{nutz2016utility} with some modifications for our setting. Let us denote, for all $\om^t \in \Om_t$ and $h_t \in \R^d$,
	$$
		V'_t(\om^t, h_t)
		:=
		\inf_{h \in \R^d} \sup_{\theta_t \in \Lambda_{\rm int}(t,\om^{t})} \sup_{\Pb \in \Pcb^{\delta}_{\mathrm{int}}(t,\omb)}
			\log \E^{\Pb}
			\Big[
				\exp\big( g_{t+1} + h (X_{t+1} - X_t) + h_t X_t \big)
			\Big].
	$$
	By Remark \ref{rem:g_prime}, we can employ the same
	minimax theorem argument as in Lemma \ref{lemm:utili_max_dual_T1} above and
	 obtain that
	\b*
		V'_t(\om^{t}, h_t) = g'_t( \om^{t}, h_t)
		~>~ -\infty,
		~~~\mbox{if}~\NA(\Pct^{\delta}_{\mathrm{int}}(t,\om)) ~\mbox{holds true}.
	\e*
	 In view of {\rm (iii)} in Remark \ref{rem: NA Pintdelta omega}, this holds true outside a $\Pc$-polar set $N$.
	
	Further, let us denote by $\Uc(\Om_t \times \R^d)$ the universal $\sigma$-field on $\Om_t \times \R^d$.
	Notice that $g_{t+1}$ is assumed to be upper semi-analytic,
	the graph set $\big[ \big[ \Qcb_{0}^{*}(t) \big] \big]$ is analytic by Lemma \ref{qanalytic}, $(\omb^t, \Qb) \in \Omb_t \times \Bf(\Omb_{t+1}) \mapsto \Ec \big(\Qb, \Pcb^{\delta}_{\mathrm{int}}(t,\omb^t) \big)$ is lower semi-analytic by \cite[Lemma 4.2] {bartl2016exponential} and \cite[Proposition 7.47]{BertsekasShreve.78}.
	It  follows from the measurable selection argument (see e.g.~\cite[Propositions 7.26, 7.47, 7.48]{BertsekasShreve.78}) that
	the maps $\omb^{t} \mapsto g_t(\omb^{t})$ is upper semi-analytic.
	 As $\big[ \big[\mathrm{int}K^*_t\big] \big]$ is Borel and hence $\big[ \big[\Lambda_{\rm int}(t, \cdot)\big] \big]$ is also Borel, it follows from \cite[Proposition 7.47]{BertsekasShreve.78} that $(\om^{t}, h_t) \mapsto g'_t(\om^{t}, h_t)$ is upper semi-analytic, hence belongs to $\Uc(\Om_t \times \R^d)$.
	

	\vspace{0.5em}
		
	Next, we claim that the function
	$$
	    \phi(\om^t, h_t, h):~=~ \sup_{\theta_t \in \Lambda_{\rm int}(t,\om^{t})} \sup_{\Pb \in \Pcb^{\delta}_{\mathrm{int}}(t,\omb)}
			\log \E^{\Pb}
			\Big[
				\exp\big( g_{t+1} + h (X_{t+1} - X_t) + h_t X_t \big)
			\Big].
	$$
	is $\Uc(\Om_t \times \R^d) \otimes \Bc(\R^d)$-measurable. To see this, we first fix $h$ and $h_t$. Then from the same argument as above, as $\left[ \left[ \Pcb^{\delta}_{\mathrm{int}}(t) \right] \right]$ is analytic and by \cite[Propositions 7.26, 7.47, 7.48]{BertsekasShreve.78},
	we have that $$(\om^t, \theta_t) \mapsto \sup_{\Pb \in \Pcb^{\delta}_{\mathrm{int}}(t,\omb)}\log \E^{\Pb}[\exp\big( g_{t+1} + h (X_{t+1} - X_t) + h_t X_t)]$$
is upper semi-analytic.
	Again, $\big[ \big[\mathrm{int}K^*_t\big] \big]$ is Borel 
	implies that $\big[ \big[\Lambda_{\rm int}(t, \cdot)\big] \big]$ is also Borel, by \cite[Proposition 7.47]{BertsekasShreve.78}, we have that $\om^t \mapsto \phi(\om^t, h_t, h)$ is upper semi-analytic.
	On the other hand, for fixed $\om^t$, it follows by an application of Fatou's lemma (as \cite[Lemma 4.6]{bartl2016exponential})
	that $(h, h_t) \mapsto \phi(\om^t, h_t, h)$ is lower semi-continuous.
 Moreover, as $(h, h_t) \mapsto \phi(\om^t, h_t, h)$ is convex, by \cite[Lemma 4.5]{BC}, we have that $\phi$ is indeed $\Fc_t \otimes \Bc(\R^d) \otimes \Bc(\R^d)$-measurable, and thus belongs to $\Uc(\Om_t \times \R^d) \otimes \Bc(\R^d)$.

	Let us consider the random set $$\Phi(\om^t, h_t):=\left\{ h \in \R^d : \phi(\om^t, h_t, h)=g'_t(\om^t, h_t) \right\}.$$ By the previous arguments,  we have that $\left[ \left[ \Phi \right] \right]$ is in
 $\Uc(\Om_t \times \R^d) \otimes \Bc(\R^d)$. Thus $\Phi$ admits an $\Uc(\Om_t \times \R^d)$-measurable selector $h_{t+1}$ on the universally measurable set $\Phi(\om^t, h_t) \neq \emptyset$; cf. the corollary and scholium of \cite[Theorem 5.5]{Leese}. Moreover, Lemma \ref{lemm:utili_max_dual_T1} and Remark \ref{rem: NA Pintdelta omega} imply that $\Phi(\om^t, h_t) \neq \emptyset$ holds true outside a $\Pc$-polar set $N$, it yields that $h_{t+1}$ solves the infimum $\Pc$-q.s.
	\qed

\paragraph{The multi-period case: the final proof}

	We provide a last technical lemma and then finish the proof of Proposition \ref{prop:enlarged_utility_max_duality}.
	Recall that $g_{t+1}:= \Omb_{t+1} \to \R \cup \{ \infty\}$ is a given upper semi-analytic functional,
	such that
	$g_{t+1}( \om^{t+1}, \theta_0, \cdots, \theta_{t+1})$ depends only on $(\om^{t+1}, \theta_{t+1})$,
	and $g_t$ is defined in \eqref{eq:def_g_n}.
	Given a universally measurable random variable $Y_t: \Omb_t \rightarrow \R_+$, we define
	$$
		 \Qcb_{Y_t,t}^*:=\{ \Qb \in \Qcb_0|_{\Omb_t}: \E^{\Qb}\big[ g_{t}^- \big]+\Ec \big(\Qb, \Pcb_{\mathrm{int}}|_{\Omb_t} \big) < +\infty, ~\E^{\Qb}[Y_t] < + \infty \}.
	$$

	\begin{Lemma} \label{lemm:concatenateQ}
		Let $t+1 \le T$, then for any universally measurable random variable $Y_{t+1}: \Omb_{t+1} \to \R_+$ and $\eps > 0$,
		there is a universally measurable random variable $Y^{\eps}_t: \Omb_t \to \R_+$ such that
		\be \label{eq:DualConcatenation1}
			\sup_{\Qb \in \Qcb_{Y^{\eps}_t,t}^*} \Big\{ \E^{\Qb} [ g_t ] - \Ec \big(\Qb, \Pcb_{\mathrm{int}}|_{\Omb_t} \big) \Big\}
			\le
			\sup_{\Qb \in \Qcb_{Y_{t+1},t+1}^*} \Big\{ \E^{\Qb} [ g_{t+1} ] - \Ec \big(\Qb,  \Pcb_{\mathrm{int}}|_{\Omb_{t+1}} \big) \Big\} + \eps.
		\ee
	\end{Lemma}
	\proof{}  
	$\mathrm{(i)}$
	In view of Corollary \ref{coro:phiBetter},
	we can assume w.l.o.g. that $Y_{t+1} \equiv 0$.
	Then Lemma \ref{qanalytic} and a measurable selection argument (see e.g. Proposition 7.50 of \cite {BertsekasShreve.78}) guarantees that
	there exists a universally measurable kernel $\Qb_{t}^{\eps}(\cdot): \Omb_t \rightarrow \Bf(\Omb_1)$ such that $\delta_{\omb} \otimes\Qb_{t}^{\eps}(\omb) \in  \Qcb_{0}^{*}(t,\omb)$ for all $\omb \in \Omb_t$, and
	$$
		g_t(\omb)
		~\le~
		\E^{\delta_{\omb} \otimes\Qb_{t}^{\eps}(\omb)}[ g_{t+1}]
		-
		\Ec \big(\Qb_{t}^{\eps}(\omb), \Pcb_{\mathrm{int}}(t,\omb) \big)
		+
		\eps.
	$$
	
	\noindent $\mathrm{(ii)}$
	Let us define $Y^{\eps}_t$ by
	$$
		Y_t^{\eps}(\cdot)
		:=
		\E^{\delta_{\cdot} \otimes \Qb_{t}^{\eps}(\cdot)}\big[  g_{t+1}^- +|X_{t+1}-X_t| \big] + \Ec \big(\Qb_{t}^{\eps}(\cdot),\Pcb_{\mathrm{int}}(t,\cdot) \big).
	$$
	By the definition of $\Qcb_{0}^{*}(t, \cdot)$ and \cite[Lemma 4.2]{bartl2016exponential},
	$Y_t^{\eps}$ is $\R_+$-valued and universally measurable.

	Then for any $\Qb \in \Qcb_{Y_t^{\eps},t}^*$, one has
	\b*
		&& \E^{\Qb \otimes \Qb_{t}^{\eps}(\cdot)}\left[g_{t+1}^- + |X_{t+1}-X_t|\right] +\Ec(\Qb \otimes \Qb_{t}^{\eps}(\cdot), \Pcb_{\mathrm{int}}|_{\Omb_{t+1}}) \\
		&=& \E^{\Qb}\left[ \E^{\Qb_{t}^{\eps}(\cdot)}[g_{t+1}^- +|X_{t+1}-X_t| ] \right] + \Ec(\Qb, \Pcb_{\mathrm{int}}|_{\Omb_{t}}) + \E^{\Qb}\left[\Ec(\Qb_{t}^{\eps}(\cdot), \Pcb_{\mathrm{int}}(t, \cdot))\right] \\
		&\le& \Ec(\Qb, \Pcb_{\mathrm{int}}|_{\Omb_{t}}) + \E^{\Qb}[Y_{t}^{\eps}] < + \infty,
	\e*	
	where the first equality follows from Lemma 4.4 of \cite{bartl2016exponential}. Further, $\Qb \otimes \Qb_{t}^{\eps}(\cdot)$ is a martingale measure on $\Omb_{t+1}$ by the martingale property of $\Qb$ and $\Qb_{t}^{\eps}(\cdot)$.
	Finally, because $\Qb \lll \Pcb_{\mathrm{int}}|_{\Omb_{t}}$ and $\Qb_{t}^{\eps}(\cdot) \lll \Pcb_{\mathrm{int}}(t,\cdot)$,
	it follows that $\Qb \otimes \Qb_{t}^{\eps}(\cdot) \lll \Pcb_{\mathrm{int}}|_{\Omb_{t+1}}$.
	This implies that $\Qb \otimes \Qb_{t}^{\eps}(\cdot) \in \Qcb_{0,t+1}^*$.

	Thus for any $\Qb \in \Qcb_{Y_t^{\eps},t}^*$, one has
	\b*
	 \E^{\Qb}[g_t] - \Ec(\Qb, \Pcb_{\mathrm{int}}|_{\Omb_t})
	 &\leq& \E^{\Qb} \left[\E^{\Qb_{t}^{\eps}(\cdot)}[ g_{t+1}] - \Ec \big(\Qb_{t}^{\eps}(\cdot), \Pcb_{\mathrm{int}}(t,\cdot) \big) + \eps \right] - \Ec(\Qb, \Pcb_{\mathrm{int}}|_{\Omb_t}) \\
	 &=& \E^{\Qb \otimes \Qb_{t}^{\eps}(\cdot)} [g_{t+1}] - \Ec(\Qb \otimes \Qb_{t}^{\eps}(\cdot), \Pcb_{\mathrm{int}}|_{\Omb_{t+1}}) + \eps \\
	 &\leq& \sup_{\Qb \in \Qcb_{0,t+1}^*} \left\{ \E^{\Qb} [ g_{t+1} ] - \Ec(\Qb, \Pcb_{\mathrm{int}}|_{\Omb_{t+1}}) \right\} +\eps.
	\e*
	We hence conclude the proof as $\Qb \in \Qcb_{Y_t^{\eps},t}^*$ is arbitrary.
	\qed

	\vspace{0.5em}

	\noindent {\bf Proof of  Proposition \ref{prop:enlarged_utility_max_duality}.}
	We will use an induction argument.
	First, Proposition \ref{prop:enlarged_utility_max_duality} in case $T=1$ is already proved in Lemma \ref{lemm:utili_max_dual_T1}.
	Next, assume that Proposition \ref{prop:enlarged_utility_max_duality} holds true for the case $T = t$,
	we then consider the case $T = t+1$.

	\vspace{1mm}

	In the case $T = t+1$, let us denote $g_{t+1} := g := \xi \cdot X_{t+1}$.
	It is clear that $g_{t+1}$ is a Borel random variable and
	$g_{t+1}(\om^{t+1}, \theta_0, \cdots, \theta_{t+1})$ depends only on $(\om^{t+1}, \theta_{t+1})$.
	Let $g_t$ be defined by \eqref{eq:def_g_n}.
	Since Proposition \ref{prop:enlarged_utility_max_duality} is assumed to hold true for the case $T = t$,
	it follows that there is $\hat H = (\hat H_1, \cdots, \hat H_t) \in \Hc_t$ such that,
	for any universally measurable random variable $Y_t : \Omb_t \to \R_+$, one has
	\be \label{eq:case_t}
		\sup_{\Pb \in \Pcb_{\mathrm{int}}} \log \E^{\Pb} \big[ \exp( g_t + (\hat H \circ X)_t \big) \big]
		&=&
		\sup_{\Qb \in \Qcb_{Y_t, t}^*} \Big\{ \E^{\Qb} [ g_t ] - \Ec \big(\Qb, \Pcb_{\mathrm{int}}|_{\Omb_t} \big) \Big \}.
	\ee
	Then with the function $h_{t+1}$ defined in Lemma \ref{lemm:utility_max_selec_h}, we define
	\be \label{eq:def_Hh_t1}
		\hat H_{t+1}(\om^t) ~:=~ h_{t+1}( \om^t, \hat H_t(\om^{t-1})).
	\ee
		
	Further, for any $\Pb \in \Pcb_{\mathrm{int}}$, one has the representation $\Pb = \Pb_0 \otimes \cdots \otimes \Pb_t$,
	where $\Pb_s(\cdot)$ is measurable kernel in $\Pcb^{\delta}_{\mathrm{int}}(s,\cdot)$.
	It follows by direct computation that
	\b*
		\E^{\Pb} \left[ \exp\left( g_{t+1} + (\hat H \circ X)_{t+1} \right) \right]
		\!\!\!&=&\!\!\!
		\E^{\Pb_0 \otimes \cdots \otimes \Pb_{t-1}} \Big[ \exp \Big(
			\log \E^{\Pb_t}
			\Big[
				\exp\big( g_{t+1} +  (\hat H \circ X)_{t+1} \big)
			\Big]  \Big) \Big] \\
		\!\!\!&\le&\!\!\!
		\E^{\Pb} \Big[ \exp \Big(  \sup_{\Pb' \in \Pcb_{\mathrm{int}}^{\delta}(t,\cdot) }
			\log \E^{\Pb'}
			\Big[
				\exp\big( g_{t+1} +  (\hat H \circ X)_{t+1} \big)
			\Big]  \Big) \Big] \\
		\!\!\!&\le&\!\!\!
		\sup_{\Pb \in \Pcb_{\mathrm{int}}} \E^{\Pb} \big[ \exp \big( g'_t(\cdot, \hat H_t) + (\hat H \circ X)_{t-1} -\hat H_t X_{t-1} \big) \big],	 
	\e*
	where the last inequality follows by the definition of $\hat H_{t+1}$ in \eqref{eq:def_Hh_t1} and Lemma \ref{lemm:utility_max_selec_h}.
	Taking the supremum over $\Pb \in \Pcb_{\mathrm{int}}$,
	it follows from the definition of $g'_t$ in \eqref{eq:def_g_t_p} together with a dynamic programming argument that
	\be \label{eq:DPP_hedging}
		\sup_{\Pb \in \Pcb_{\mathrm{int}}} \E^{\Pb} \left[ \exp\left( g_{t+1} + (\hat H \circ X)_{t+1} \right) \right]
		\!\!\!&\le&\!\!\!
		\sup_{\Pb \in \Pcb_{\mathrm{int}}} \E^{\Pb} \big[ \exp \big( g_t +\hat H_t X_t + (\hat H \circ X)_{t-1} -\hat H_t X_{t-1} \big) \big] \nonumber \\
		\!\!\!&=&\!\!\!
		 \sup_{\Pb \in \Pcb_{\mathrm{int}}} \E^{\Pb} \big[ \exp \big( g_t + (\hat H \circ X)_t \big) \big].
	\ee
	 Then for any universally measurable random variable $\varphi: \Omb \to \R_+$,
	we set $Y_{t+1} := \varphi$ and use sequentially Lemma \ref{lemm:concatenateQ}, \eqref{eq:case_t}, \eqref{eq:DPP_hedging},
	to obtain
	\b*
		\sup_{\Qb \in \Qcb^*_{\varphi}} \Big\{ \E^{\Qb} [ g_{t+1} ] - \Ec(\Qb, \Pcb_{\mathrm{int}}|_{\Omb_{t+1}}) \Big\}
		&\ge&
		\sup_{\Qb \in \Qcb_{Y_t, t}^*} \Big\{ \E^{\Qb} [ g_t ] - \Ec \big(\Qb, \Pcb_{\mathrm{int}}|_{\Omb_t} \big) \Big \}\\
		&=&
		\sup_{\Pb \in \Pcb_{\mathrm{int}}} \log \E^{\Pb} \big[ \exp( g_t + (\hat H \circ X)_t \big) \big]\\
		&\ge&
		\sup_{\Pb \in \Pcb_{\mathrm{int}}} \log \E^{\Pb} \left[ \exp\left( g_{t+1} + (\hat H \circ X)_{t+1} \right) \right] \\
		&\ge&
		\inf_{H \in \Hc} \sup_{\Pb \in \Pcb_{\mathrm{int}}} \log \E^{\Pb} \left[ \exp\left( g_{t+1} + ( H \circ X)_{t+1} \right) \right].
	\e*
	Because the reverse inequality is the weak duality in  Lemma \ref{lemm:weak_dual_utility},
	we obtain the equality everywhere in the above formula,
	which is the duality result \eqref{eq:max_util_duality_enlarg} for the case $T = t+1$.
	In particular, $(\hat H_1, \cdots, \hat H_t, \hat H_{t+1})$ is the optimal trading strategy for the case $T= t+1$.
	\qed

\subsection{Proof of Theorem \ref{thm:utility_max_duality}(Case $e \geq 1$)}
\label{subsec:proof_duality}

In this section, we are interested in the utility maximization problem with semi-static strategy.
To take into account of the transaction costs caused by trading the static options $(\zeta_i, i=1, \cdots, e)$, we work in the framework of \cite{BDT17} and introduce a further enlarged space by
	$$
		\widehat \Lambda := \prod_{i=1}^e [-c_i, c_i],
		~~
		\Omh := \Omb \x \widehat \Lambda ,
		~~
		\Fch_t := \Fcb_t \otimes \Bc\big(\widehat \Lambda \big),
		~~
		\Pch_{\mathrm{int}} := \left\{
		\Ph \in \Bf(\Omh)~: \Ph|_{\Omb} \in \Pcb_{\mathrm{int}}
		\right\},
	$$
	and define
	$$
		\hat f_i : \Omh ~\longrightarrow~ \R ,
		~~~
		\hat f_i(\omh) = \zeta_{i}(\om) \cdot X_T(\omb) - \hat \theta_i
		~~\mbox{for all}~\omh = (\omb, \hat \theta) = (\om, \theta, \hat \theta) \in \Omh.
	$$
	The process $(X_t)_{0 \le t \le T}$ and the random variable $g := \xi \cdot X_T$ defined on $\Omb$ can be naturally extended on $\Omh$.
	
	We can then consider the exponential utility maximization problem on $\Omh$:
 $$
 \inf_{(H,\ell) \in \Hc \times \R^e} \sup_{\Ph \in \Pch_{\mathrm{int}}} \log \E^{\Ph} \left[ \exp\left( g+\sum_{i=1}^e \ell_i \hat f_i + (H \circ X)_T \right) \right].
 $$
Let us also introduce
$$
\Qch^{*}_e ~:=~\left\{ \Qh \in \Bf(\Omh)  ~:~~
\begin{aligned}
& \Qh \lll \Pch_{\mathrm{int}}, ~ X ~\mbox{is}~(\Fbh, \Qh) \mbox{-martingale}, ~\E^{\Qh}[\hat f_i] = 0,~ i=1,\cdots, e, & \\
&\E^{\Qh}\big[ (\xi \cdot X_T)_- \big] +\Ec(\Qh, \Pch_{\mathrm{int}}) < + \infty &
\end{aligned}
\right\},
$$
and
$$
\Qch^{*}_{e,\varphi}:= \{ \Qh \in \Qch^{*}_e : \E^{\Qh}[\varphi] < + \infty \},~~\mathrm{for ~all}~\varphi:\Omh \to \R_+.
$$
It is easy to employ similar arguments for Lemma \ref{lemm:equiv_entropy} and Proposition \ref{prop:reformulation} to obtain
\b*
&&\inf_{(\ell,\eta) \in \Ac_e}
			\sup_{\P \in \Pc}
			\log \E^{\P} \left[ \exp \left(\left(\xi  - \sum_{i=1}^{e}   \left(\ell_i  \zeta_i - |\ell_{i}| c_{i} \1_{d} \right)  - \sum_{t=0}^T \eta_t \right)^d  \right) \right] \\
&=&  \inf_{(H,\ell) \in \Hc \times \R^e} \sup_{\Ph \in \Pch_{\mathrm{int}}} \log \E^{\Ph} \left[ \exp\big( g+\sum_{i=1}^e \ell_i \hat f_i + (H \circ X)_T \big) \right],
\e*
and
$$
\sup_{ (\Q,Z) \in \CPSt_e^*} \big\{ \E^{\Q} \big[ \xi \cdot Z_T\big] - \Ec(\Q,\Pc) \big \} = \sup_{ \Qh \in \Qch^{*}_e } \big\{ \E^{\Qh} \big[ g \big] - \Ec(\Qh,\Pch_{\mathrm{int}}) \big \},
$$
with $g:=\xi \cdot X_T$.
Hence, to conclude the proof of Theorem \ref{thm:utility_max_duality}(case $e \geq 1$), it is sufficient to prove that, for any universally measurable $\varphi: \Omh \to \R_+$, one has
\be \label{eq:dualityOmh}
\inf_{(H,\ell) \in \Hc \times \R^e} \sup_{\Ph \in \Pch_{\mathrm{int}}} \!\! \log \E^{\Ph} \left[ \exp\left( g+\sum_{i=1}^e \ell_i \hat f_i + (H \circ X)_T \right) \right]
= \sup_{ \Qh \in \Qch^{*}_{e,\varphi}} \!\! \Big\{ \E^{\Qh} \big[ g \big] - \Ec(\Qh,\Pch_{\mathrm{int}}) \Big\}.
\ee

	Let us first provide a useful lemma.
\begin{Lemma} \label{lemm:SupRepEntro}
	Let $g: \Omb \rightarrow \R$ be upper semi-analytic, and assume that $\NA2(\Pc)$ holds. Assume either that $e=0$,
		or that $e \ge 1$  and for all $\ell \in \R^{e}$ and $\eta\in \Ac$,  \eqref{eq: NA avec option et CT} holds.		
		Then, for all $\varphi: \Omh \to \R_+$, one has
	\be \label{eq:SupRepEntro}
		\inf \Big\{
			y  \in \R~: y + \sum_{i=1}^{e} \ell_i \hat{f}_i + (H \circ X)_T \ge g, ~\Pch_{\mathrm{int}} \mbox{-q.s.}, l \in \R^e, H \in \Hc
		\Big\}
		~=~
		\sup_{\Qb \in \Qch^{*}_{e,\varphi}} \E^{\Qb} \big[ g \big].
	\ee
\end{Lemma}
\proof
By Proposition \ref{prop:equiv_NA}, $\NA2(\Pc)$ implies $\NA(\Pcb_{\mathrm{int}})$.
For the case $e=0$, as observed by \cite[Lemma 3.5]{bartl2016exponential}, Lemma 3.3 of \cite{BouchardNutz.13} has indeed proved the following stronger version of fundamental lemma with $T=1$:
\be \label{eq:FundLemma}
0 \in \mathrm{ri} \{~ \E^{\Qb}[\Delta X],~ \Qb \in \Qcb^{*}_{\varphi} \}.
\ee
Using \eqref{eq:FundLemma}, we can proceed as \cite[Lemma 3.5, 3.6, Theorem 4.1]{BouchardNutz.13} to prove \eqref{eq:SupRepEntro} in the case without options($e=0$).

For the case $e \geq 1$, we can argue by induction. Suppose the super-replication theorem with $e-1$ options holds with $g=\hat f_e$:
\be \label{eq:superrepOption}
		\hat \pi_{e-1}(g)
		&:=&
		\inf\Big\{
			y ~: y + \sum_{i=1}^{e-1} \ell_i \hat f_i + (H \circ X)_T \ge g, ~\Pch_{\mathrm{int}} \mbox{-q.s.}, ~\ell \in \R^{e-1},~ H \in \Hc
		\Big\}  \nonumber   \\
		&=&  \sup_{\Qh \in \Qch^{*}_{e-1,\varphi}} \E^{\Qh}[ g],
\ee
and we shall pass to $e$. By the no arbitrage condition \eqref{eq: NA avec option et CT}, there is no $H \in \Hc$, $\ell_1, \cdots, \ell_{e-1}$ and $\ell_{e}\in \{-1,1\}$ such that $ \sum_{i=1}^{e-1} \ell_i \hat f_i  + (H \circ X)_T\ge -\ell_{e}\hat f_{e}$, $\Pch_{\mathrm{int}}$-q.s.
	It follows that $\hat \pi_{e-1}(\hat f_{e}),\hat \pi_{e-1}(-\hat f_{e})>0$, which, by \cite[Lemma 3.12]{BouchardNutz.13} and \eqref{eq:superrepOption}, implies that there is $\Qh_{-}, \Qh_+ \in \Qch_{e-1,\varphi}^{*}$ such that
	\be\label{eq: pie de fe+1 avec 0 au milieu}
		- \hat \pi_{e-1}(- \hat f_{e}) ~<~ \E^{\Qh_-}[\hat  f_{e}]
		~<~ 0 ~<~  \E^{\Qh_+}[ \hat f_{e}] ~<~ \hat \pi_{e-1}(\hat f_{e}).
	\ee
	In particular, we have
\be \label{eq:relinterior1}
0 \in \mathrm{ri} \{~ \E^{\Qh}[\hat f_e],~ \Qh \in \Qch^{*}_{e-1,\varphi} \}.
\ee
Then we can argue line by line as \cite[Proof of Theorem 3.1(case $e \geq 1$)]{BDT17} to prove that
\b*
		\mbox{there exists a sequence}~ \big(\Qh_n \big)_{n\ge 1} \subset \Qch_{e,\varphi}^{*}
		~\mbox{s.t.} ~\E^{\Qh_n}[ g] \to \hat  \pi_{e}(g),~~~
	\e*
	 which shows that
	$$
		\sup_{\Qh \in \Qch_{e,\varphi}^{*}} \E^{\Qh}[ g]
		~\ge~
		\hat \pi_{e}(g).
	$$
To conclude, it is enough to notice that the reverse inequality is the classical weak duality which can be easily obtained from \cite[Lemmas A.1 and A.2]{BouchardNutz.13}.
\qed

\vspace{0.5em}

\noindent {\bf Proof of Theorem \ref{thm:utility_max_duality} (case $e \geq 1$).}
Notice that \eqref{eq:dualityOmh} has been proved for the case $e=0$ in Section \ref{subsec:proof_e0}, although the formulations are slightly different. The proof is still based on the induction as in the proof of \cite[Theorem 2.2]{bartl2016exponential}. Let us assume that \eqref{eq:dualityOmh} holds for $e-1 \ge 0$ and then prove it for the case of $e$.
	Define
	\b*
	J: \Qch^{*}_{e-1,\varphi} \times \R \rightarrow \R, ~~~ (\Qh, \beta) \mapsto \E^{\Qh}[g] + \beta \E^{\Qh}[\hat f_e] - H(\Qh,\Pch_{\mathrm{int}}).
	\e*
Clearly, $J$ is concave in the first argument and convex in the second argument. { By \eqref{eq:relinterior1},} $J$ satisfies the compactness-type condition (14) in \cite{bartl2016exponential}, thus we can apply the minimax theorem. Using the induction hypothesis and the same arguments as in \cite{bartl2016exponential}, we have
\be \label{eq:dualWithOption}
&&\inf_{(H,\ell) \in \Hc \times \R^e} \sup_{\Ph \in \Pch_{\mathrm{int}}} \log \E^{\Ph} \left[ \exp\left( g+\sum_{i=1}^e \ell_i \hat f_i + (H \circ X)_T \right) \right] \nonumber  \\
&=& \inf_{\beta \in \R} \min_{(H,\ell) \in \Hc \times \R^{e-1}} \sup_{\Ph \in \Pch_{\mathrm{int}}} \E^{\Ph} \left[ \exp\left( g+\sum_{i=1}^{e-1} \ell_i \hat f_i + \beta \hat f_e + (H \circ X)_T \right) \right]  \nonumber  \\
&=&
	 \inf_{\beta \in \R} \sup_{\Qh \in \Qch^{*}_{e-1,\varphi}} J (\Qh, \beta) \\
&=& \sup_{\Qh \in \Qch^{*}_{e-1,\varphi}} \inf_{\beta \in \R} J (\Qh, \beta) = \sup_{\Qh \in \Qch^{*}_{e,\varphi}} \big( \E^{\Qh}[g] - H(\Qh,\Pch_{\mathrm{int}}) \big). \nonumber
\ee
The duality holds as a consequence.  Moreover, from (15) of \cite{bartl2016exponential}, $\forall c < \inf_{\beta \in \R} \sup_{\Qh \in \Qch^{*}_{e-1,\varphi}} J (\Qh, \beta)$, $\exists n, \mbox{ s.t. for all } \beta$ satisfying $|\beta|>n, \sup_{\Qh \in \Qch^{*}_{e-1,\varphi}} J (\Qh, \beta) > c$. Thus \eqref{eq:dualWithOption} can be rewritten as
$$\inf_{|\beta| \le n} \sup_{\Qh \in \Qch^{*}_{e-1,\varphi}} J (\Qh, \beta).$$ Now the lower-continuity of $\beta \mapsto \sup_{\Qh \in \Qch^{*}_{e,\varphi}} J(\Qh, \beta)$ implies the existence of some $\hat \beta$ such that
$$
\sup_{\Qh \in \Qch^{*}_{e,\varphi}} J(\Qh, \hat \beta) = \inf_{\beta \in \R} \sup_{\Qh \in \Qch^{*}_{e,\varphi}} J(\Qh, \beta).
$$
Combining $\hat \beta$ with the optimal strategy with $e-1$ options $(\hat H, \hat \ell ^{\star})$, we deduce the existence of an optimal strategy for $e$ options, namely $(\hat H, \hat \ell) := (\hat H, (\hat \ell ^{\star}, \hat \beta))$. Using the construction \eqref{eq:H2eta},
	one can obtain $(\hat \eta, \hat \ell)$ explicitly attaining the infimum in \eqref{eq:util_max_dual_reform1}
	from $(\hat H,  \hat \ell )$ which is constructed already in previous steps.
\qed

\subsection{Proof of Proposition \ref{prop:utilityindiff}} \label{sec:utilityindiff}

	Using the expression in \eqref{indifftwoequiv}, one has
	\b*
		\lim_{\gamma\rightarrow\infty}\pi_{\gamma}(\xi)
		&=&
		\lim_{\gamma\rightarrow\infty}\sup_{ (\Q,Z) \in \CPSt_e^*}\left\{ \E^{\Q} \big[ \xi \cdot Z_T\big] -\frac{1}{\gamma} \Ec(\Q,\Pc)\right\},
	\e*
	where the r.h.s. is increasing in $\gamma$.
	Replacing the limit by supremum, and then interchanging the order of two supremums, we have
	\b*
		\lim_{\gamma\rightarrow\infty}\pi_{\gamma}(\xi)
		~=~
		\sup_{ (\Q,Z) \in \CPSt_e^*}\sup_{\gamma} \left\{ \E^{\Q} \big[ \xi \cdot Z_T\big] -\frac{1}{\gamma} \Ec(\Q,\Pc)\right\}
		~=~
		\sup_{ (\Q,Z) \in \CPSt_e^*} \E^{\Q} \big[ \xi \cdot Z_T\big].
	\e*
	By similar arguments as in Section 3.2 of \cite{BDT17}, we can reformulate the problem at the r.h.s. on the enlarged space $\Omh$ and then use Lemma \ref{lemm:SupRepEntro} to obtain that
	\b*
		\sup_{ (\Q,Z) \in \CPSt_e^*} \E^{\Q} \big[ \xi \cdot Z_T\big]
		~=~
		\sup_{ \Qh \in \Qch^{*}_{e} } \E^{\Qh} \big[ \xi \cdot X_T \big]
		~=~
		\pi(\xi).
	\e*
	This concludes the proof.
	\qed

\vspace{5em}

\begin{APPENDICES}
\textbf{Appendix: Exponential utility maximization duality without transaction cost}

	In this appendix, we shall present an auxiliary result on the exponential utility maximization problem without transaction cost by applying the same procedure as in the proof of Theorem \ref{thm:utility_max_duality}.
	This allows to extend the main results in Bartl \cite{bartl2016exponential} without a restrictive $\om$-wise no-arbitrage condition.
	Moreover, an auxiliary result in the dominated case consists a key ingredient in the proof of our main result in Theorem \ref{thm:utility_max_duality} with transaction cost (in particular in Lemma \ref{lemm:concatenateQ}).

	\vspace{0.5em}

	Throughout this appendix, we stay in the context of Section \ref{subsec:prelimin},
	where $\Om := \Om_1^T$ is a (product) Polish space with the raw canonical filtration $\F^0 = (\Fc^0_t)_{0 \le t \le T}$ and the universally completed filtration $\F = (\Fc_t)_{0 \le t \le T}$ and $\Fc := \Fc_T$.
	The space $(\Om, \Fc)$ is equipped with a family of (possibly) non-dominated probability measures $\Pc$ defined by \eqref{eq:def_Pc} with a given family of classes of probability measures $\Pc_t(\om)$ on $\Om_1$, that is,
	$$
		\Pc :=
		\big\{
			\P := \P_0 \otimes \P_1 \otimes \cdots \otimes \P_{T-1} ~: \P_t(\cdot) \in \Pc_t(\cdot) \mbox{ for } t\le T-1
		\big\},
	$$
	which satisfies the measurability condition \eqref{eq:AnalyticGraph}.
	We take the $\F^0$-adapted process $(S_{t})_{0 \le t \le T}$ in \eqref{eq: S in int}
	and let it represent the discounted stock price, which can be traded without any transaction cost.
	Finally, by a slight abuse of language, we denote $g: \Om \rightarrow \R$ an upper semi-analytic random variable representing the payoff of a derivative option, and
	let
	$$
		\Hc
		~:=~
		\{ \mbox{All}~ \F \mbox{-predictable processes} \}
	$$
	represent the set of all admissible trading strategies,
	and denote $(H\circ S)_T := \sum_{t=1}^T H_t \cdot (S_{t+1} - S_t)$.
	Following Bouchard and Nutz \cite{BouchardNutz.13}, we define the quasi-sure no-arbitrage condition $\NA(\Pc)$ by
	\be
		(H \circ S)_T \ge 0, ~\Pc \mbox{-q.s.} ~~\Longrightarrow~~(H \circ S)_T = 0, ~\Pc \mbox{-q.s.}~~ \mbox{~for all~} H \in \Hc.
	\ee
	Further, for each $t=0, \cdots, T-1$ and $\om^t \in \Om_t$, we define the no-arbitrage condition $\NA(\Pc_t(\om^t))$ by
	\be \label{eq:def_noarb_1P}
		h \cdot \Delta S_{t+1}(\om^t, \cdot) \ge 0, ~\Pc_t(\om^t) \mbox{-q.s.}
		~\Longrightarrow~
		h \cdot \Delta S_{t+1}(\om^t, \cdot) = 0, ~\Pc_t(\om^t) \mbox{-q.s.}~
		\mbox{~for all~} h \in \R^d.
	\ee
	Recall also that (by Lemma 4.6 of \cite{BouchardNutz.13}) the set $N_t=\{ \om^t \in \Om_t: \NA(\Pc_t(\om^t)) ~\mbox{fails} \}$ is $\Pc$-polar if $\NA(\Pc)$ holds.

	\vspace{0.5em}
	
	Let us denote by $\Qc_0$ the collection of measures $\Q \in \Bf(\Om)$ such that $\Q \lll \Pc$ and $S$ is an $(\F, \Q)$-martingale.
	Then given a universally measurable random variable $\varphi : \Om \rightarrow \R_+$, we define
	$$
		\Qc^*_0 ~:= \big\{ \Q \in \Qc_0 ~: \E^{\Q}\big[ g^- \big]+\Ec(\Q,\Pc) < \infty \big\}
		\mbox{~and~}
		\Qc^*_{\varphi}:=\{ \Q \in \Qc^*_0: \E^{\Q}[\varphi] < + \infty \}.
	$$

	\begin{Theorem} \label{thm:Bartlimprove}
		Let $g: \Om \to (-\infty, + \infty]$ be upper semi-analytic and suppose that $\NA(\Pc)$ holds.
		Then for any universally measurable $\varphi : \Om \rightarrow \R_+$, one has
		$$
			V
			:=
			\inf_{H \in \Hc} \sup_{\P \in \Pc} \log \E^{\P} \big[ \exp\big( g + (H \circ S)_T \big) \big]
			=
			\sup_{\Q \in \Qc^*_{\varphi}} \big\{ \E^{\Q}[g]-\Ec(\Q, \Pc) \big\}.
		$$
		Moreover, the infimum over $H \in \Hc$ is attained by some optimal trading strategy $\hat{H}$.
	\end{Theorem}

	\begin{Remark}
		In Bartl \cite{bartl2016exponential}, the above { result} is proved under the condition that $\NA(\Pc_t(\om^t))$ holds for all $t=0, \cdots, T-1$ and all $\om^t \in \Om_t$.
		As explained { in Remark 2.5 of \cite{bartl2016exponential}}, the main reason to use this $\om$-wise no-arbitrage condition (rather than the quasi-sure no-arbitrage condition $\NA(\Pc)$) is the measurability issue due to their dynamic programming procedure.
		Our alternative procedure allows to overcome this measurability difficulty.
	\end{Remark}

\section{Some technical lemmas}
         In this section, we shall give some technical lemmas which will be used in both Section \ref{subsec:dominatedUtility} and Section \ref{subsec:nondominatedUtility}. Firstly, by using the same arguments as in Lemma \ref{lemm:weak_dual_utility},
one obtains the next weak duality.
	\begin{Lemma} \label{lemm:WeakDualNoTransac}
		Under the same conditions as Theorem \ref{thm:Bartlimprove}, one has
		$$
			\inf_{H \in \Hc} \sup_{\P \in \Pc} \log \E^{\P} \big[ \exp\big( g + (H \circ S)_T \big) \big]
			~\geq~
			\sup_{\Q \in \Qc^{*}_0} \big\{ \E^{\Q}[g]-\Ec(\Q, \Pc) \big\}.
		$$
	\end{Lemma}

	Next, for all $t \in \{ 0, \cdots, T -1\}$,
	we consider an upper semi-analytic function $g_{t+1}: \Om_{t+1} \to \R \cup \{ \infty \}$,
	and define
	\b*
		g_t(\om^t)
		~:=
		\sup_{\Q \in \Qc_{0}^{*}(t,\om^t)}
		\Big\{
			\E^{\Q}[ g_{t+1}] - \Ec \big(\Q, \Pc_t(\om^t) \big)
		\Big\}, ~~\mbox{for all}~ \om^t \in \Om_t,
	\e*
	where
	\b*
		\Qc_{0}^{*}(t,\om^t)
		&\!\!:=\!\!&
		\Big\{
			\delta_{\om^t} \otimes \Q \in \Bf(\Om_{t+1})  ~:
		~\Q \lll \Pc_t(\om^t),~\E^{\Q} \big[ S_{t+1}(\om^t, \cdot) - S_t(\om^t) \big] = 0, \\
		&&~~~~~~~~~~~~~~~~~~~~~
		\E^{\Q}\big[ g_{t+1}^-(\om^t, \cdot) + |S_{t+1}(\om^t, \cdot)-S_t(\om^t)| \big]+\Ec \big(\Q, \Pc_t(\om^t) \big) < \infty
		\Big\}.
	\e*
	Further, given a universally measurable random variable $Y_{t+1} : \Om_{t+1} \rightarrow \R_+$, we introduce
	$$
		\Qc_{Y_{t+1}}^{*}(t,\om^t)
		~:=~
		\big\{
			\Q \in \Qc_{0}^{*}(t,\om^t)
			~:~
			\E^{\Q}[ Y_{t+1}(\om^t, \cdot) ] < \infty
		\big\}.
	$$
	{ Moreover}, for any universally measurable random variable $Y_t: \Om_t \rightarrow \R_+$, we denote
	$$
		\Qc_{Y_t,t}^{*}:=\{ \Q \in \Qc_0|_{\Om_t}: \E^{\Q}[g_t^-] + \Ec(\Q, \Pc)< + \infty,~ \E^{\Q}[Y_t] < + \infty \}.
	$$

	\begin{Lemma}	\label{lemm:QcYrelated}	
		For any universally measurable random variable $Y_{t+1}: \Om_{t+1} \to \R_+$, one has
		\b*
			g_t(\om^t)
			~=
			\sup_{\Q \in \Qc_{Y_{t+1}}^{*}(t,\om^t)}
			\Big\{
			\E^{\Q}[ g_{t+1}] - \Ec \big(\Q, \Pc_t(\om^t) \big)
			\Big\}, ~~\Pc\mbox{-q.s}.
		\e*
		{ In addition, if }$Y_{t+1}$ is assumed to be Borel measurable, the graph set
		$$
			\big[ \big[ \Qc_{Y_{t+1}}^{*}(t) \big] \big]
			~:=~
			\big\{ (\om, \Q): \om \in \Om_t,
				\Q \in \Qc_{Y_{t+1}}^{*}(t,\om)
			\big\}
			~\text{is analytic.}
		$$	
	\end{Lemma}
	\proof
	The first result { is consequent on} the one-period duality result in Theorem 3.1 of Bartl \cite{bartl2016exponential}
	(see also our Remark \ref{rmk:independentY}),
	and the second result follows essentially the same arguments as in the proof of Lemma \ref{qanalytic}.
	\qed

	\begin{Lemma} \label{lemm:utility_max_selec_h2}
		Assume that $\NA(\Pc)$ holds true.
		Then $g_t$ is upper semi-analytic, and there exists a universally measurable map $h_{t+1}: \Om_t \to \R^d$, together with a $\Pc$-polar set $N \subset \Om_t$ such that, for all $\om \in N^c$, one has
		$$
			g_t(\om^t)
			~=~
			\sup_{\P \in \Pc_t(\om^t)} \log \E^{\P}
			\Big[
				\exp\big( g_{t+1}(\om^t, \cdot) +  h_{t+1}(\om^t) (S_{t+1} (\om^t, \cdot) - S_t(\om^t)) \big)
			\Big]
                         >
                         - \infty.
		$$	
	\end{Lemma}
	\proof
	The argument is similar to Lemma \ref{lemm:utility_max_selec_h}, so we shall provide here a sketch of the proof.
  As $g_{t+1}$ is upper semi-analytic, $\big[ \big[ \Qc_{0}^{*}(t) \big] \big]$ is analytic from Lemma \ref{lemm:QcYrelated} and $(\om^t,\Q) \in \Om_t \times \Bf(\Om_{1}) \mapsto -\Ec(\Q, \Pc_t(\om^t))$ is upper semi-analytic by \cite[Lemma 4.2]{bartl2016exponential} and \cite[Proposition 7.47]{BertsekasShreve.78},
 it follows from Lemma \ref{lemm:QcYrelated} and a measurable selection argument(see e.g. \cite[Proposition 7.26, 7.47, 7.48]{BertsekasShreve.78}) that $\om^t \mapsto g_t$ is upper semi-analytic. By defining
$$
V^*_t (\om^t):= \inf_{h_{t+1} \in \R^d} \sup_{\P \in \Pc_t(\om^t)}   \log \E^{\P}
			\Big[
				\exp\big( g_{t+1}(\om^t, \cdot) + h_{t+1} (S_{t+1}(\om^t, \cdot) - S_t(\om^t)) \big)
			\Big],
$$
and applying Theorem 3.1 of \cite{bartl2016exponential}, we obtain that
$$
			g_t(\om^t)
			~=~
			V^*_t (\om^t)
                         >
                         - \infty,~~~\mbox{if}~\NA(\Pc_t(\om^t)) ~\mbox{holds true}.
		$$
		As $\NA(\Pc)$ holds, this is valid outside a $\Pc$-polar set $N$.

By defining $\phi_t(\om^t, h_{t+1}):= \sup_{\P \in \Pc_t(\om^t)} \log \E^{\P}
			\Big[
				\exp\big( g_{t+1}(\om^t, \cdot) + h_{t+1} (S_{t+1}(\om^t, \cdot) - S_t (\om^t)) \big)
			\Big]$, we can argue similarly as Lemma \ref{lemm:utility_max_selec_h} to see that $(\om^t, h_{t+1}) \mapsto \phi_t$ is in $\Fc_t \otimes \Bc(\R^d)$. Let us now consider the random set
			$$\Phi(\om^t):= \{ h \in \R^d: \phi(\om^t,h)= g_t(\om^t) \}.$$
	The previous arguments yield that $\left[ \left[ \Phi \right] \right]$ is in $\Fc_t \otimes \Bc(\R^d)$. Thus by Lemma 4.11 of \cite{BouchardNutz.13}, $\Phi$ admits an $\Fc_t$-measurable selector $h_{t+1}$ on the universally measurable set $\Phi(\om^t) \neq \emptyset$. { Moreover, Theorem 3.1 of \cite{bartl2016exponential} implies that $\Phi(\om^t) \neq \emptyset$ holds true outside a $\Pc$-polar set $N$, thus $h_{t+1}$ solves the infimum $\Pc$-q.s.
\qed}

\section{Proof of Theorem \ref{thm:Bartlimprove} in a dominated case}
\label{subsec:dominatedUtility}

	We first provide the proof of  Theorem \ref{thm:Bartlimprove} in a dominated case,
	where $\Pc$ is a singleton, i.e.  $\Pc= \{\P\}$, for $\P = \P_0 \otimes \P_1 \otimes \cdots \otimes \P_{T-1}$,
	where $\P_t(\om^t) \in \Pc_t(\om^t)$ for all $\om^t \in \Om^t$ and all $t\le T-1$.
	In particular, $\P_t: \Om_t \to \Bf(\Om_{t+1})$ is a regular conditional probability distribution(r.c.p.d.) of $\P$ knowing $\Fc_t^0$.
	We can assume without loss of generality that $\Pc_t(\om^t) = \{\P_t(\om^t) \}$.
	Moreover, let $\Fc^{\P}_t$ denote the $\P$-completion of the $\sigma$-field $\Fc_t$,
	then any $\Fc^{\P}_t$-measurable random variable can be modified to a Borel measurable random variable in sense of $\P$-a.s.
	
	\vspace{0.5em}
	
	The following lemma is an analogue of Lemma \ref{lemm:concatenateQ},
	and in this dominated context, the measurability issue is much easier to treat.

\begin{Lemma} \label{lemma:dominated_concatenation}
	Assume the same conditions in Theorem \ref{thm:Bartlimprove} and that $\Pc = \{\P\}$.
	Then for all $t \le T-1$ and all $\Fc^{\P}_{t+1}$-measurable random variable $Y_{t+1}: \Om_{t+1} \to \R_+$,
	there is a Borel random variable $Y_t: \Om_t \to \R_+$ such that
	\be \label{eq:DualConcatenation1}
		\sup_{\Q \in \Qc_{Y_t,t}^{*}} \Big\{ \E^{\Q} [ g_t ] - \Ec(\Q, \P|_{\Om_t}) \Big\}
		~\le~
		\sup_{\Q \in \Qc_{Y_{t+1},t+1}^{*}} \Big\{ \E^{\Q} [ g_{t+1} ] - \Ec(\Q, \P|_{\Om_{t+1}}) \Big\}.
	\ee
\end{Lemma}
	\proof
	Under the reference probability $\P$, for any { $\Fc^{\P}_{t+1}$-measurable} random variable $Y_{t+1}$, there exists a Borel measurable random variable $Y^*_{t+1}$, such that $Y_{t+1}=Y^*_{t+1}, \P\mbox{-a.s.}$ and thus $\E^{\Q}[Y_{t+1}]=\E^{\Q}[Y^*_{t+1}], \mbox{ for all } \Q \in \Qc_{Y_{t+1},t+1}^{*}$. So we can assume w.l.o.g. that $Y_{t+1}$ is Borel measurable. By Lemma \ref{lemm:QcYrelated} together with a measurable selection argument (see e.g. Proposition 7.50 of \cite {BertsekasShreve.78}), for any $\eps > 0$,
	there exists a universally measurable kernel $\Q_{t}^{\eps}(\cdot): \Om_t \rightarrow \Bf(\Om_1)$ such that $\delta_{\om} \otimes\Q_{t}^{\eps}(\om) \in  \Qc_{Y_{t+1}}^{*}(t,\om)$ for all $\om \in \Om_t$, and
	$$
		g_t(\om)
		~\le~
		\E^{\delta_{\om} \otimes\Q_{t}^{\eps}(\om)}[g_{t+1}]
		-
		\Ec \big( \Q_{t}^{\eps}(\om), \P_t(\om) \big)
		+
		\eps.
	$$
	The rest arguments are almost the same as in Step $\mathrm{(ii)}$ of the proof of Lemma \ref{lemm:concatenateQ} and we shall omit the details.
	\qed

\vspace{0.5em}

\noindent {\bf Proof of Theorem \ref{thm:Bartlimprove} when $\Pc = \{\P\}$}.
	 We can argue by induction as in the proof of Proposition \ref{prop:enlarged_utility_max_duality}. Noticing $\NA(\{\P\})$ holds, the case $T=1$ is proved by Theorem 3.1 of \cite{bartl2016exponential}.
Suppose the case $T=t$ is verified with optimal strategy $\hat H:=(\hat H_1, \cdots, \hat H_{t})$:
	\be
		 \log \E^{\P} \big[ \exp(g_t + (\hat H \circ S)_t \big) \big]
		&=&
		\sup_{\Q \in \Qc_{Y_t, t}^{*}} \Big\{ \E^{\Q} [g_t ] - \Ec \big(\Q, \P|_{\Om_t} \big) \Big \},
	\ee
	and we shall pass to the $T=t+1$ case. Denoting $g_{t+1} := g$, defining $\hat H_{t+1}(\om^t): = h_{t+1}$ as in Lemma \ref{lemm:utility_max_selec_h2}, setting $Y_{t+1} := \varphi$ for any universally measurable random variable $\varphi: \Om \to \R_+$, and letting $Y_t$ be given as in Lemma \ref{lemma:dominated_concatenation},
	it follows that
	\b*
		&&
		\sup_{\Q \in \Qc^{*}_{\varphi}} \Big\{ \E^{\Q} [g_{t+1} ] - \Ec(\Q, \P|_{\Om_{t+1}}) \Big\}
		~\ge~
		\sup_{\Q \in \Qc_{Y_t, t}^{*}} \Big\{ \E^{\Q} [g_t ] - \Ec \big(\Q, \P|_{\Om_{t}} \big) \Big \}\\
		&=&
		\log \E^{\P} \big[ \exp(g_t + (\hat H \circ S)_t \big) \big]\\
		&=&
		\log \E^{\P_0 \otimes \cdots \otimes \P_{t-1}} \Big[ \exp \Big(
			\log \E^{\P_t}
			\Big[
				\exp\big(g_{t+1} +  (\hat H \circ S)_{t+1} \big)
			\Big]  \Big) \Big] \\
		&=&
		\log \E^{\P} \left[ \exp\left(g_{t+1} + (\hat H \circ S)_{t+1} \right) \right] \\
		&\ge&
		\inf_{H \in \Hc} \log \E^{\P} \left[ \exp\left(g_{t+1} + ( H \circ S)_{t+1} \right) \right],
	\e*
	where the first inequality follows by Lemma \ref{lemma:dominated_concatenation} and the third line follows by Lemma \ref{lemm:utility_max_selec_h2}. As the reverse inequality is the weak duality by Lemma \ref{lemm:WeakDualNoTransac}, we have proved the case $T=t+1$. In particular, $(\hat H_1, \cdots, \hat H_t, \hat H_{t+1})$ is the optimal trading strategy for the case $T= t+1$.
	\qed

	\vspace{3mm}

The following corollary  serves as an important technical step in the proof of Lemma \ref{lemm:concatenateQ}.

\begin{Corollary} \label{coro:phiBetter}
	Assume the same conditions in Theorem \ref{thm:Bartlimprove} and let $\P \in \Pc$ be fixed.
	Then for any universally measurable random variables
	$g: \Om \to \R$ and $\varphi: \Om \to \R_+$, and any $\Q^* \in \Qc_0^*$,
	one has
        \be \label{eq:phiBetter}
        		\E^{\Q^*} [ g ] - \Ec \big(\Q^*,  \P \big)
        		~\leq~
        		\sup_{\Q \in \Qc_{\varphi}^{*}} \Big\{ \E^{\Q} [  g] - \Ec \big(\Q,  \P \big)  \Big\}.
        \ee
\end{Corollary}
	\proof Without loss of generality, we can assume that $\Ec(\Q^*, \P) < \infty$.
	In this case, one has $\Q^* \ll \P$ and the classical no-arbitrage condition $\NA(\{\Q^*\})$ holds.
	Let us denote
	\b*
		\Qc^{**}_{\varphi}:= \{ \Q \in \Qc^*_{\varphi}: \Ec(\Q, \Q^*) < + \infty \},
	\e*
	Using the weak duality of Lemma \ref{lemm:WeakDualNoTransac} and applying Theorem \ref{thm:Bartlimprove}
	({ in the context of the fixed} probability space $(\Om, \Fc, \Q^*)$),
	we have
	\b*
		&&
		\E^{\Q^*} [ g ] - \Ec \big(\Q^*,  \P \big)
		~~=~~
		\E^{\Q^*} \left[ g - \log \frac{d\Q^*}{d\P}\right] - \Ec(\Q^*,\Q^*) \\
		&\le&
		\inf_{H \in \Hc} \log \E^{\Q^*} \left[ \exp\left( g - \log \frac{d\Q^*}{d\P} + (H \circ X)_T \right) \right] \\
		&=&
		\sup_{\Q \in \Qc_{\varphi}^{**}} \left( \E^{\Q} \left[ g - \log \frac{d\Q^*}{d\P}\right] - \Ec \left(\Q,  \Q^* \right) \right) \\
		&\leq&
		\sup_{\Q \in \Qc_{\varphi}^{*}} \left( \E^{\Q} \left[ g - \log \frac{d\Q^*}{d\P}\right] - \Ec \left(\Q,  \Q^* \right) \right)
		~~=~~
		\sup_{\Q \in \Qc_{\varphi}^{*}} \left( \E^{\Q} [ g ] - \Ec \left(\Q,  \P \right) \right).
	\e*
\qed

\section{Proof of Theorem  \ref{thm:Bartlimprove}} \label{subsec:nondominatedUtility}

	We now provide the proof of Theorem  \ref{thm:Bartlimprove} in the general (possibly) non-dominated case.

\begin{Lemma} \label{lemm:concatenateQ2}
	Let $t+1 \le T$, then for any universally measurable random variable $Y_{t+1}: \Om_{t+1} \to \R_+$ and $\eps > 0$,
	there is a universally measurable random variable $Y^{\eps}_t: \Om_t \to \R_+$ such that
	\be \label{eq:DualConcatenation1}
		\sup_{\Q \in \Qc_{Y^{\eps}_t,t}^*} \Big\{ \E^{\Q} [ g_t ] - \Ec \big(\Q, \Pc|_{\Om_t} \big) \Big\}
		\le
		\sup_{\Q \in \Qc_{Y_{t+1},t+1}^*} \Big\{ \E^{\Q} [ g_{t+1} ] - \Ec \big(\Q,  \Pc|_{\Om_{t+1}} \big) \Big\} + \eps.
	\ee
\end{Lemma}
\proof
 In view of Corollary \ref{coro:phiBetter},
	we can assume w.l.o.g. that $Y_{t+1} \equiv 0$.
	By Lemma \ref{lemm:QcYrelated} and a measurable selection argument (see e.g. Proposition 7.50 of \cite {BertsekasShreve.78}), for any $\eps > 0$,
	there exists a universally measurable kernel $\Q_{t}^{\eps}(\cdot): \Om_t \rightarrow \Bf(\Om_1)$ such that $\delta_{\om} \otimes\Q_{t}^{\eps}(\om) \in  \Qc_{Y_{t+1}}^{*}(t,\om)$ for all $\om \in \Om_t$, and
	$$
		g_t(\om)
		~\le~
		\E^{\delta_{\om} \otimes\Q_{t}^{\eps}(\om)}[g_{t+1}]
		-
		\Ec \big(\Q_{t}^{\eps}(\om), \Pc_t(\om) \big)
		+
		\eps.
	$$
	The rest argument is similar to Step $\mathrm{(ii)}$ of the proof of Lemma \ref{lemm:concatenateQ}, and we omit it here.
\qed

\vspace{0.5em}

\noindent \textbf{Proof of Theorem \ref{thm:Bartlimprove}.}
	Notice that under $\NA(\Pc)$, the results in case $T=1$ follows from Theorem 3.1 of \cite{bartl2016exponential}.
	Suppose that when $T=t$ the duality holds with optimal strategy $(\hat H_1, \cdots, \hat H_t)$.
	Denoting $g_{t+1} := g$, defining $\hat H_{t+1}(\om^t): = h_{t+1}$ as in Lemma \ref{lemm:utility_max_selec_h2},
	setting $Y_{t+1} := \varphi$ and letting $Y_t$ be given in Lemma \ref{lemm:concatenateQ2},
	we apply similar argument as in the dominated context in Section \ref{subsec:dominatedUtility} to obtain
	\b*
		\sup_{\Q \in \Qc^*_{\varphi}} \Big\{ \E^{\Q} [ g_{t+1} ] - \Ec(\Q, \Pc|_{\Om_{t+1}}) \Big\}
		&\ge&
		\sup_{\Q \in \Qc_{Y_t, t}^*} \Big\{ \E^{\Q} [ g_t ] - \Ec \big(\Q, \Pc|_{\Om_t} \big) \Big \}\\
		&=&
		\sup_{\P \in \Pc} \log \E^{\P} \big[ \exp( g_t + (\hat H \circ S)_t \big) \big]\\
		&\ge&
		\sup_{\P \in \Pc} \log \E^{\P} \left[ \exp\left( g_{t+1} + (\hat H \circ S)_{t+1} \right) \right] \\
		&\ge&
		\inf_{H \in \Hc} \sup_{\P \in \Pc} \log \E^{\P} \left[ \exp\left( g_{t+1} + ( H \circ S)_{t+1} \right) \right].
	\e*
	This together with Lemma \ref{lemm:WeakDualNoTransac} implies the duality result.
	Moreover,  $(\hat H_1, \cdots, \hat H_{t+1})$ is the optimal trading strategy for the case $T = t+1$.
\qed

 \end{APPENDICES}

%


%
%
%

\section*{Acknowledgments.}
X. Tan gratefully acknowledges the financial support of the ERC 321111 Rofirm, the ANR Isotace, and the Chairs Financial Risks and Finance and Sustainable Development.
	His work has also benefited from the financial support of the Initiative de Recherche ``M\'ethodes non-lin\'eaires pour la gestion des risques financiers'' sponsored by AXA Research Fund.
	X. Yu is partially supported by the Hong Kong Early Career Scheme under grant no. 25302116 and the Hong Kong Polytechnic University central research grant under no.15304317.


\bibliographystyle{informs2014} 
\bibliography{ExpUtility_TC18} 


\end{document}